\documentclass[prb, twocolumn, superscriptaddress]{revtex4}  
\usepackage{amssymb}
\usepackage{amsmath}
\usepackage{bbm}
\usepackage{graphicx}
\usepackage{color}

\begin{document}

\author{Philipp Werner}
\affiliation{Department of Physics, University of Fribourg, 1700 Fribourg, Switzerland}
\author{Martin Eckstein}
\affiliation{Institute of Theoretical Physics, University of Hamburg, 20355 Hamburg, Germany}
\author{Naoto Tsuji}
\affiliation{Department of Physics, University of Tokyo, Hongo, Tokyo 113-0033, Japan}
\affiliation{RIKEN Center for Emergent Matter Science (CEMS), Wako, Saitama 351-0198, Japan}

\title{Nonequilibrium DMFT approach to time-resolved Raman spectroscopy}

\date{\today}

\begin{abstract}
Raman spectroscopy uses light scattering to extract information on low-energy excitations of solids. The Raman process is described by diagrams which are fourth order in the light-matter interaction, and in particular the resonant contribution, which involves four different space-time arguments, is difficult to evaluate. 
If one instead simulates explicitly  the incoming (classical) light pulse, the Raman signal is given by the  outgoing photon flux and can be determined from a  two-point correlation function. Such a formalism can be used to compute the time-resolved Raman spectrum of non-equilibrium systems, as well as nonlinear signals which are higher order in the incoming field, such as hyper Raman scattering. Here we explain how to implement this time-dependent formalism within the dynamical mean field theory framework. The method is illustrated with applications to the Holstein-Hubbard model in the strong electron-phonon coupling regime. We demonstrate hyper Raman scattering in measurements with strong probe fields and frequency mixing signals in the presence of a pump field, and simulate the evolution of Stokes and anti-Stokes features after photo-excitations of metallic and Mott insulating systems.
\end{abstract}

\maketitle

\section{Introduction}

Ultrafast optical spectroscopies provide detailed information on the nonequilibrium evolution of photo-excited materials.\cite{Giannetti2016,Torre2021} They allow to study single-particle and collective excitations, order parameters and phase transitions, as well as the coupled dynamics of spin, orbital and lattice degrees of freedom. Photoemission spectroscopy provides information on the occupied electronic states of the solid,\cite{Sobota2021} (x-ray) absorption spectroscopy on particle-hole excitations and unoccupied states, \cite{Stamm2010,Baykusheva2022,Lojewski2023} while Raman scattering\cite{Yang2017} and resonant inelastic x-ray scattering (RIXS) \cite{Dean2016,Wang2020,Ament2011} probe various types of low-energy excitations.\cite{Devereaux2007}

An advantage of light-scattering experiments such as Raman and RIXS is the freedom to target specific excitations by selecting the polarization of both the incident and scattered light. This is one of the reasons why Raman scattering has been used extensively to study cuprates, where the physics in the nodal and antinodal region differs substantially.\cite{Devereaux2007,Saichu2009} Moreover, the comparison of scattering with energy loss $\omega$ (Stokes signal $I(\omega)$) and with energy gain $\omega$ (anti-Stokes signal $I(-\omega)$) allows to obtain not only information on the spectrum of the low energy modes, but also on their occupation. This makes time-resolved Raman spectroscopy a potentially powerful tool for probing the relaxation pathways of various degrees of freedom in photo-excited solids. For example, the anti-Stokes signal has been used to measure phonon temperatures during photo-induced structural phase transitions \cite{Fausti2009} and the lifetime of phonons in graphene,\cite{Kang2010,Yan2009} and theoretical proposals exist for the transient thermometry of electronic or magnetic degrees of freedom. \cite{Matveev2019,Wang2018}

In diagrammatic analyses of electronic Raman scattering, we distinguish (i) contributions which involve excitations into electronic intermediate states via the linear light-matter interaction (``$j\cdot A$"), and (ii) contributions which arise from the diamagnetic light-matter interaction. We refer to these two types as resonant and nonresonant diagrams, respectively, because the contributions (i) become dominant when the electronic intermediate state is (near) resonant to the probe.   

For a single clean  band of Bloch electrons, resonant contributions are suppressed, because  an optical transition with momentum transfer $q=0$ does not couple the ground state and electronically excited states. In interacting electron systems, however, electronic excitations are possible at $q=0$, such as excitations between the Hubbard bands in a Mott insulator, or excitations between phonon sidebands in an electron-phonon coupled system. In this case, resonant contributions to the Raman signal may be relevant for the interpretation of experiments. For example, Raman thermometry often assumes that the ratio $I(-\omega)/I(\omega)$ of Stokes and anti-Stokes signals in thermal equilibrium is a universal function of temperature, given by the Boltzmann factor $ e^{-\beta\omega}$. However, this universality relies on a symmetry of the Raman tensor which may no longer hold when near-resonant intermediate states are involved in the scattering (see, e.g., the discussion in Ref.~\onlinecite{Walter2020}).  For electronic Raman scattering, the anti-Stokes/Stokes ratio takes its universal form when one restricts the analysis to off-resonant contributions,\cite{Matveev2019,Wang2018,Freericks2001,Shvaika2004,Shvaika2005,Matveev2010} but for the resonant contributions, the ratio may depend on microscopic details of the scattering process. To unlock the full potential of Raman spectroscopy  for characterizing the non-equilibrium evolution of correlated electrons, a theoretical framework which can treat both resonant and nonresonant Raman contributions is therefore desirable. 

The theoretical calculation of resonant contributions to light scattering (Raman or RIXS) is challenging already in equilibrium.\cite{Devereaux2007,Ament2011} In nonequilibrium situations the problem becomes even harder,\cite{Matveev2019,Wang2018,Chen2019}  because the standard diagrammatic calculations require the evaluation of a four-point correlation function. An alternative way to calculate spectroscopic signals in time-dependent situations is to explicitly simulate the probe pulse as a classical light field, and take the limit of a weak probe field in the numerical results. For example, this has been exploited to compute time-resolved optical conductivities,\cite{Shao2016, Werner2019} and to simulate scattering within wave-function based approaches.\cite{Zawadzki2020,Rincon2021} Recently, such a formalism was developed to calculate RIXS, wherein the first half of the RIXS process -- the excitations of core electrons to the conduction states -- is simulated explicitly.\cite{Eckstein2021,Werner2021a} In this approach, the diagrammatic calculation only needs to deal with the photon emission part of the scattering process and involves a two-point correlation function. This method has been implemented within a dynamical mean field theory (DMFT)\cite{Georges1996} framework -- first proposed in the equilibrium context in Refs.~\onlinecite{Hariki2018,Hariki2020} -- and applied to single-orbital \cite{Eckstein2021} and multi-orbital Hubbard systems,\cite{Werner2021a} electron-phonon systems,\cite{Werner2021b} and a model for rare earth nickelates.\cite{Werner2023}   

Here, we introduce a related two-point formalism for the scattering to compute the resonant contributions to the time-resolved Raman scattering signal, using nonequilibrium DMFT.\cite{Aoki2014} (The nonresonant diagrams can be obtained by a separate two point correlation function.) While no core levels are involved in this case, the Raman process also corresponds to the excitation of electrons into virtual or valence states, and their relaxation back into the low-energy manifold via the emission of photons. We simulate the excitation part of this process by treating the incoming light as a classical field. By measuring an appropriate two-point correlation function, the resonant contributions to the Raman amplitude can be evaluated for each outgoing photon frequency by a post-processing scheme. 

In both the two-point formalism and the four-point formalism, the full diagrammatic evaluation of the respective two- and four-point functions in terms of Green's functions requires vertex corrections. Even if vertex corrections are not fully included, the  two-point point formalism has a number of technical advantages which will be discussed in this manuscript (Sec.~\ref{sec_4vs2point}). Most obviously, the two-point formalism is not restricted to weak probe pulse amplitudes, and is thus suitable for the study of higher-order nonlinearities in the signal. As an example we discuss the emergence of hyper Raman signals at high probe amplitudes, which can be understood as Raman scattering from a Floquet dressed band structure. 

The paper is organized as follows. In Sec.~\ref{sec:method} we describe the methodology. The adaptation of the conventional DMFT formalism for nonresonant Raman scattering to simulations with explicit Raman drive is presented in an appendix and is found to produce results which are qualitatively similar to the resonant signal for high enough frequency of the probe light.  Section~\ref{sec:results} presents test results for the resonant Raman signal of the Hubbard-Holstein model in and out of equilibrium, while Sec.~\ref{sec:conclusions} is a short conclusion.  

\begin{figure}[t]
\begin{center}
\includegraphics[angle=0, width=\columnwidth]{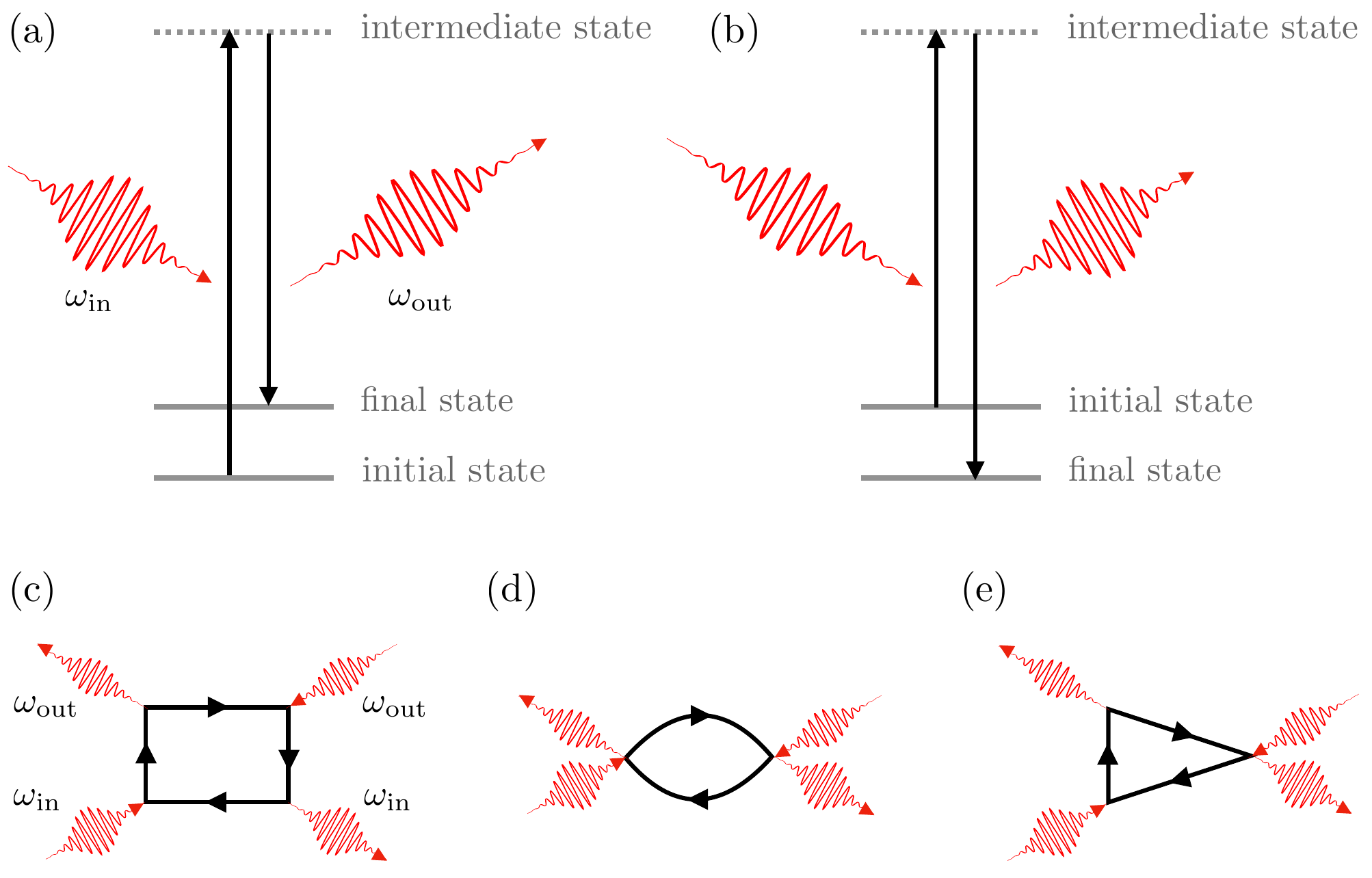}
\caption{
(a,b) Illustration of the light scattering (Raman) process. 
Panel (a) shows a Stokes process and (b) an anti-Stokes process.
Lower panels: simple diagrams (without vertex corrections) for 
(c) resonant Raman scattering, 
(d) nonresonant Raman scattering, and
(e) mixed scattering.
}
\label{fig_illustration}
\end{center}
\end{figure}

\section{Formalism}
\label{sec:method}

\subsection{Light scattering processes}

The light scattering process is schematically illustrated in Fig.~\ref{fig_illustration}. Electrons are excited by a laser pulse with frequency $\omega_\text{in}$ and an envelope $s_\text{in}(t)$ into virtual (nonresonant Raman scattering) or long-lived (resonant Raman scattering) intermediate states. The decay of these electrons back into the low-energy manifold of states leads to the emission of photons with frequency $\omega_\text{out}$. In panel (a), the final state energy of the system is higher than the initial state energy, which corresponds to a Stokes process. At nonzero initial temperature or in photoexcited systems, the final state energy can also be lower than the initial energy, see panel (b), which corresponds to an anti-Stokes process. 

\subsection{Model and coupling to the field}

We will demonstrate the formalism by considering a Holstein-Hubbard model on an infinite-dimensional hypercubic lattice. The lattice Hamiltonian is given by 
\begin{align}
H=&\sum_{ij\sigma} v_{ij}(t)c^\dagger_{i\sigma}c_{j\sigma}+U\sum_i n_{i\uparrow}n_{i\downarrow}-\mu\sum_{i\sigma}n_{i\sigma}\nonumber\\
&+\sum_{i}\sqrt{2}g(n_{i\uparrow}+n_{i\downarrow}-1)X_i + \frac{\omega_0}{2}\sum_i (X_i^2+P_i^2),\label{eq_H}
\end{align}
where $c_{i\sigma}$ denotes the annihilation operator for an electron on site $i$ with spin $\sigma$, $X_i$ ($P_i$) the phonon position (momentum) operator, $\omega_0$ the phonon frequency, $g$ the electron-phonon coupling, $U$ the local Hubbard interaction and $\mu$ the chemical potential. The light-matter coupling is described by the Peierls phase of the hopping amplitude,\cite{Peierls1933,Aoki2014}
\begin{equation}
v_{ij}(t)=t_{ij}e^{-i\phi_{ij}},
\end{equation}
where $t_{ij}$ is the hopping amplitude from site $j$ to site $i$ in the absence of a field. Assuming a long wavelength compared to the lattice spacing, the phase is 
\begin{align}
\label{peierlks}
\phi_{ij}= \frac{q}{\hbar}\vec{A}\cdot \vec{r}_{ij}= \frac{qa}{\hbar}\vec{A}\cdot \vec{\eta}_{ij},
\end{align}
with ${\vec A}(t)$ the vector potential, $q$ the electron charge, and ${\vec r}_{ij}={\vec R}_i-{\vec R}_j$ the vector connecting the sites $j$ and $i$. In the second equality, we have expressed the lattice vector as $\vec{r}_{ij} = a \vec{\eta}_{ij}$ with the lattice constant $a$, so that $\vec{\eta}_{ij}$ is dimensionless. We have dropped the $\vec{r}$-dependence of $\vec{A}$ since we consider the long wavelength limit.  

The incoming probe field will later be described by a classical time-dependent electric field
\begin{align}
\vec{E}(t) = \mathcal{E}_\text{in} \hat \epsilon_\text{in} \sin(\omega_\text{in}t) s(t),
\end{align}
where $\mathcal{E}_\text{in}$ is the electric field amplitude, $\hat \epsilon_\text{in}$ the incoming polarization vector, $\omega_\text{in}$ the frequency of the probe, and  $s(t)$ a dimensionless probe envelope function.  The vector potential is related to the electric field ${\vec E}(t)$ by ${\vec E}(t)=-\partial_t {\vec A}(t)$ and can be written as
\begin{align}
\vec{A}(t) = \frac{\mathcal{E}_\text{in}}{\omega_\text{in}} \hat \epsilon_\text{in} S(t),
\end{align}
where we have factored out the center frequency $\omega_\text{in}$ of the Raman pulse, so that the time-dependent function $S(t)$, defined by $\sin(\omega_\text{in}t)  s(t)=-\dot S /\omega_\text{in}$, is again dimensionless. With this, the classical field implies a Peierls phase
\begin{align}
\label{callsicalpeierls}
\phi^\text{cl}_{ij} &=A_0 (\hat e_\text{in}\cdot \vec{\eta}_{ij})S(t),\\
A_0 &=\frac{aq\mathcal{E}_\text{in}}{\hbar \omega_\text{in}}.\label{eq_A0}
\end{align}
For simulations of  time-resolved Raman spectroscopy, $s(t)$ will be of finite duration, centered around a  given probe time.  To simulate a continuous probe, $s(t)$ will be smoothly switched on (starting from $t=0$) to a constant value $s=1$.  Explicit forms of $s(t)$ are given in Sec.~\ref{sec:results}.

To describe the outgoing photon field at the quantum level, we use the representation of the electromagnetic field in a quantization volume $V$, 
\begin{align}
\vec{A}(\vec{r})&=\sum_{\vec{k},\alpha} \hat \epsilon_{\vec{k},\alpha} e^{i\vec{k}\cdot\vec{r}}\sqrt{\frac{\hbar}{2\varepsilon_r\varepsilon_0V\omega_k}}(a_{\vec{k},\alpha} + a_{-\vec{k},\alpha}^\dagger),\label{A_quantum}
\\
\vec{E}(\vec{r})&=\sum_{\vec{k},\alpha} \hat \epsilon_{\vec{k},\alpha} e^{i\vec{k}\cdot\vec{r}}\sqrt{\frac{\hbar\omega_k}{2\varepsilon_r\varepsilon_0V}}i(a_{\vec{k},\alpha} - a_{-\vec{k},\alpha}^\dagger).
\end{align}
Here $\hat \epsilon_{\vec{k},\alpha}$ is the polarization vector, which is chosen real and satisfies $\hat \epsilon_{\vec{k},\alpha}=\hat \epsilon_{-\vec{k},\alpha}$,  $\varepsilon_0$ is the vacuum dielectric constant, and $\varepsilon_r$ the  background dielectric constant. The light-matter coupling via the Peierls phase~\eqref{peierlks} can also be used for the quantum field.\cite{Li2020} The contribution to the Peierls phase for a given mode  $\gamma$ (defined by its wave-vector $\vec{k}$ and  a polarization direction), is therefore given by
\begin{align}
\label{phigamma}
\phi_{ij}^{\gamma} &=g_\gamma (\hat \epsilon_{\gamma}\cdot \vec{\eta}_{ij}) \frac{a_{\gamma}+a_{\gamma}^\dagger}{\sqrt{2}},
\end{align}
with a  dimensionless light-matter coupling constant 
\begin{align}
g_\gamma =\frac{aq}{\sqrt{\hbar V} }\frac{1}{\sqrt{\omega_\gamma \varepsilon_0\varepsilon_r}},\label{eq_g}
\end{align}
where we dropped again the spatial dependence $e^{i\vec{k}\cdot \vec{r}}$.

\subsection{Raman signal: General considerations}

We aim to compute the time-resolved or steady state Raman signal for a given outgoing mode $\gamma=``\text{out}"$. In the subsequent section, we will derive expressions for the photon generation rate $\Gamma_\text{out}(t)=\frac{d}{dt} \langle n_\text{out}(t) \rangle$ into this specific mode to leading (quadratic) order in the light-matter coupling $g_{\text{out}}$, i.e., we will compute the normalized rate
\begin{align}
\tilde \Gamma_\text{out}(t)= \lim_{g_\text{out}\to 0} \frac{\Gamma_\text{out}(t)}{A_0^2 g_\text{out}^2}.
\label{Gamma_out}
\end{align}
For a continuous probe, the Raman signal is then given by the photon flux at the detector in the steady state limit long after the switch-on of the probe, and is therefore  proportional to
\begin{align}
\label{gcw}
\tilde \Gamma_\text{out}^{\rm cw}=\lim_{t\to\infty} \tilde \Gamma_\text{out}(t).
\end{align}
In a time-resolved Raman measurement, the signal is proportional to the total photon number produced by the probe pulse,
\begin{align}
\label{ipulse}
\Delta \tilde N_\text{out}= \int_0^\infty dt \,\tilde \Gamma_\text{out}(t).
\end{align}
Note that both Eq.~\eqref{gcw} and \eqref{ipulse} still depend on the amplitude $A_0$; the normalization with $A_0^2$ in Eq.~\eqref{Gamma_out} ensures that the limit $A_0\to0$ gives the usual Raman signal, while deviations for larger pulse amplitudes correspond to higher order nonlinear scattering processes such as the hyper Raman signal. 
In Sec.~\ref{sec:results}  we will display  the normalized signals \eqref{gcw} and \eqref{ipulse}, or simply $\tilde \Gamma_\text{out}(t_\text{max})$ measured at the maximum simulation time $t_\text{max}$.

For completeness, let us also state the relation to the absolute photon count: For a continuous probe we define the Raman signal $\mathcal{R}^{\rm cw}_{\text{out}}$ as  the differential  photon flux per frequency interval $d\omega$ and solid angle $d\Omega$ around the frequency $\omega_k$ and direction $\hat k$ of the mode ``out''.  With the  rate \eqref{Gamma_out} per mode, the photon flux at the detector is given by
\begin{align}
\frac{d\sigma(t)}{d\Omega d\omega} = \tilde\Gamma_{\text{out}}(t) g_{\text{out}}^2 A_0^2\frac{dN_{k}}{d\Omega d\omega},
\end{align}
where  $dN_k$ is the number of modes per polarization in the phase space volume $d\Omega d\omega$. For a quantization box of volume $V$, $d^3k$ contains $dN_k = d^3k V/(2\pi)^3$ modes, and with $d^3k = k^2 d\Omega dk = \omega_k^2 c^{-3}  d\Omega d\omega$ we have $\frac{dN_{k}}{d\Omega d\omega}=\omega_k^2 c^{-3}  V/(2\pi)^3$. Using Eq.~\eqref{eq_g} for $g_{\text{out}}$, the Raman signal then becomes
\begin{align}
\label{raman_cw}
\mathcal{R}^{\rm cw}_{\text{out}} &= A_0^2\mathcal{R}_0  \tilde \Gamma_\text{out}^{\rm cw},
\end{align}
with the unit
\begin{align}
\mathcal{R}_0=\frac{q^2}{a\varepsilon_0\varepsilon_r} \frac{a^3}{c^3}\frac{\omega_k}{\hbar} \frac{1}{(2\pi)^3}.
\end{align}
For the time-resolved Raman measurement we can define the signal $\mathcal{R}^{\text{pulse}}_{\text{out}}$ as the total photon count  per frequency interval $d\omega$ and solid angle $d\Omega$,  accumulated from one probe pulse. Analogous to Eq.~\eqref{raman_cw}, this is given by
\begin{align}
\label{raman_pulse}
\mathcal{R}^{\text{pulse}}_{\text{out}} &= A_0^2\mathcal{R}_0  \Delta \tilde N_\text{out} 
\end{align}
in terms of the normalized signal \eqref{ipulse}.

\subsection{Calculation of the resonant Raman diagram}

To derive the diagrams for Raman scattering, we first expand $v_{ij}(t)$ up to fourth order in the light-matter coupling term:
\begin{equation}
v_{ij}=t_{ij}\Big(1-i\phi_{ij}-\frac{1}{2}\phi_{ij}^2+\frac{i}{6}\phi_{ij}^3+\frac{1}{24}\phi_{ij}^4+\ldots\Big). \label{eq_expansion}
\end{equation}
To get the normalized signal \eqref{Gamma_out}, we need to compute the total number of emitted photons at time $t$, $\langle a^\dagger_\gamma a_\gamma\rangle(t)$.
We can formally start from a quantum description of both the ingoing and outgoing modes. While general results  for nonzero temperature will be presented in the next section, it is illustrating to first derive the diagrammatic expression for a system which is initially in a state $|\psi_0^\text{el}\rangle|\psi_0^{\rm in}\rangle$ for the electrons and photons, where the first ket describes the ground state of the Holstein-Hubbard model and the second ket the state of the incoming photon field. We assume that the outgoing mode is initially unoccupied, $\langle \psi_0^{\rm in}| a^\dagger_\gamma a_\gamma|\psi_0^{\rm in}\rangle=0$. 

\begin{figure}[t]
\begin{center}
\includegraphics[angle=0, width=\columnwidth]{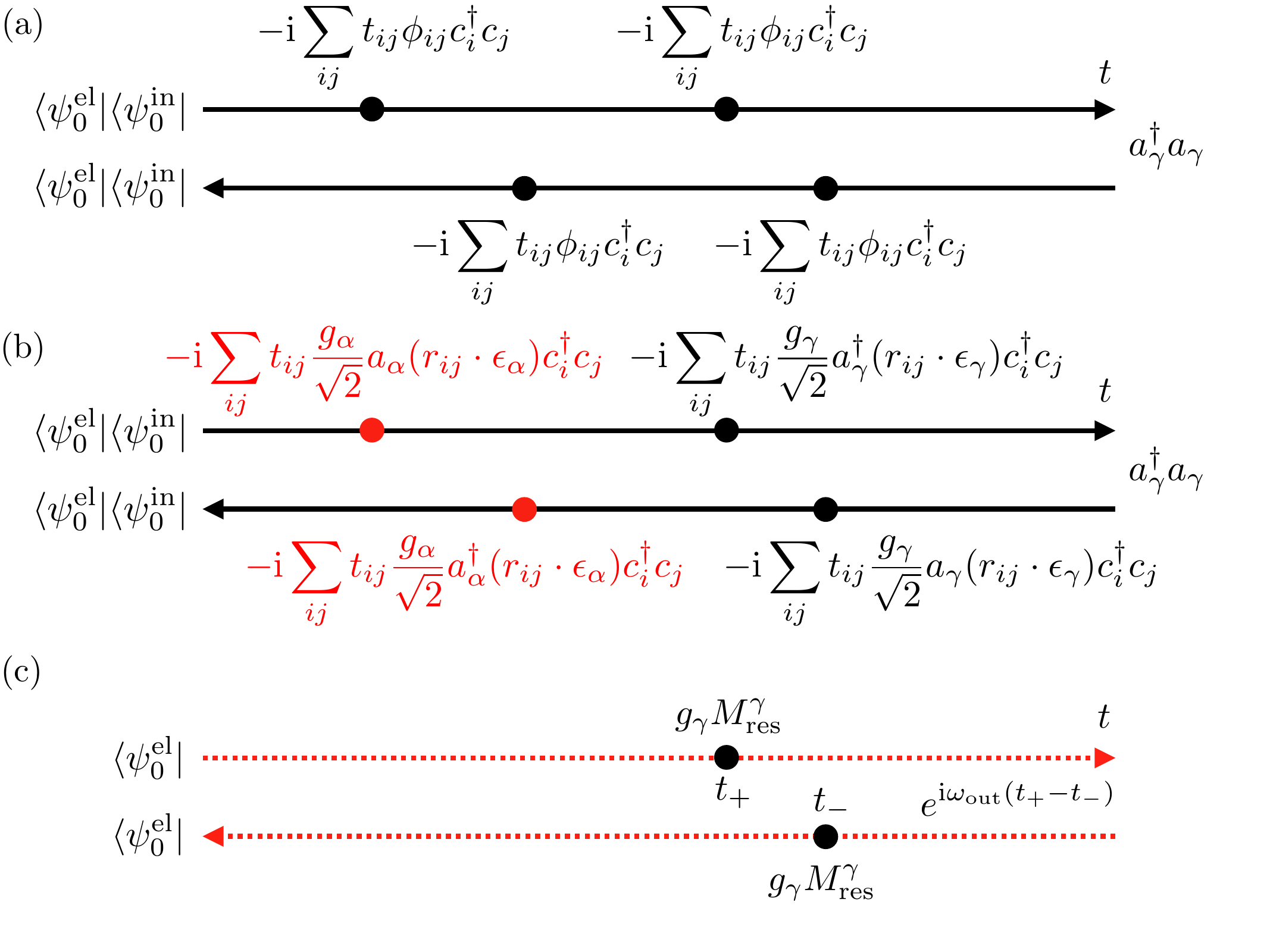}
\caption{
Derivation of the expression for resonant Raman scattering. Panel (a): Fourth order expansion in the $\phi_{ij}$ term. Solid arrows represent the time evolution without light-matter coupling, but in the presence of the driving laser field, which populates photon states with index $\alpha\ne\gamma$, where $\gamma$ is the flavor of the outgoing photon. In panel (b), we keep only those photon operators which give a nonzero contribution to the fourth order process. In panel (c), the driving laser and its effect on the system (red operators) has been replaced by a classical field which is contained in the time evolution operators (dashed arrows). The evaluation of the photon expectation value yields the factor $e^{i\omega_\gamma(t_+-t_-)}=e^{i\omega_\text{out}(t_+-t_-)}$.  
}
\label{fig_illustration_resonant}
\end{center}
\end{figure}   

The derivation is then based on standard time-dependent perturbation theory to fourth order in $\phi$ on the Kadanoff-Baym contour.\cite{Aoki2014} The terms  in the time-dependent perturbative expansion can be represented by contour diagrams as shown in Fig.~\ref{fig_illustration_resonant}. In the main text of this paper, we focus on the diagram corresponding to resonant Raman scattering, which in the conventional formalism requires the evaluation of a four-point correlation function.\cite{Devereaux2007} Within the time-dependent perturbation theory, this diagram represents a process where on the forward time contour a photon with frequency $\omega_\text{in}>0$ is absorbed from the laser and a photon with frequency $\omega_\text{out}>0$ is emitted (the order of the two processes does not matter). Then the emitted photon is measured at time $t$. On the backward branch of the time contour, the absorption of a photon with frequency $\omega_\text{out}$ and the emission of a photon with frequency $\omega_\text{in}$ occur.

In Fig.~\ref{fig_illustration_resonant}, panel (a) shows the forward and backward branches of the complex time contour and the four interaction vertices corresponding to the first-order light-matter interaction term $\propto \phi_{ij}$. The operator measured at time $t$ is the photon number $a^\dagger_\gamma a_\gamma=n_\gamma$, where $\gamma$ corresponds to the outgoing photon with frequency $\omega_\text{out}$ and polarization $\hat \epsilon_\text{out}$. In panel (b), we only show the operators which contribute to the expectation value $\langle a^\dagger_\gamma a_\gamma\rangle_0$, namely the absorption (emission) of a photon with flavor $\alpha$ on the forward (backward) branch and the emission (absorption) of a photon with flavor $\gamma$ on the forward (backward) branch. Here, $\alpha=``\text{in}"$ corresponds to the incoming light with frequency $\omega_\text{in}$ and polarization $\epsilon_\text{in}$. In the step from panel (b) to panel (c), we replace the photons of the incoming light by a classical field, corresponding to a coherent state. The effect of the classical field is explicitly taken into account to all orders in the time evolution of the system, via the classical Peierls phase \eqref{callsicalpeierls}. We have furthermore evaluated the photon expectation value, which leads to a factor $e^{\text{i}\omega_\text{out}(t_+-t_-)}$, where $t_+$ ($t_-$) denotes the times associated with the photon emission (absorption) with flavor $\gamma$ on the forward (backward) branch. 

Because the sum over the lattice sites of $r_{ij}$ times the hopping $t_{ij}c^\dagger_i c_j$ is equivalent to the sum over the velocities $v_k=\partial_k \varepsilon_k$ (with $\varepsilon_k$ the dispersion of the lattice), the vertices can be transformed into sums over momenta,  
\begin{align}
M_\text{res}^{\gamma}&=\sum_k M^{\gamma}_k n_k,\label{eq_Mres}\\
M^{\gamma}_k&=\frac{1}{\sqrt{2}}\sum_{\mu}\frac{\partial \varepsilon_k}{\partial{k_\mu}}\epsilon_{\gamma,\mu},
\end{align} 
where $n_k=\sum_{\sigma} c_{k\sigma}^\dagger c_{k\sigma}$.
 The dispersion and velocity are assumed to be spin-independent.
We thus obtain the following expression for the resonant Raman signal with frequency $\omega_\gamma=\omega_\text{out}$ and polarization $\epsilon_\gamma=\epsilon_\text{out}$:
\begin{align}
&\langle n_\text{res}^{\gamma}\rangle 
(t)=
i g_\gamma^2
 \int_{-\infty}^t \!\!\!dt_+ dt_- e^{\text{i}\omega_\text{out}(t_+-t_-)}\Pi^<_\text{res}(t_+,t_-),
\label{eq_raman}
\end{align}
where 
\begin{align}
&\Pi_\text{res}(t,t') = -i\langle \mathcal{T}_\mathcal{C} M^\gamma_\text{res}(t)M^\gamma_\text{res}(t')\rangle_\text{con}\label{eq_conn}
\end{align}
is the connected correlation function. (Note that Eq.~\eqref{eq_raman} is real, because  $\Pi^<_\text{res}(t,t')=-\Pi^<_\text{res}(t',t)^*$.) We do not consider the disconnected part of the correlation function, since this contribution corresponds to forward scattering into the direction of the main laser beam, or into the reflected signal. Equations \eqref{eq_raman} and \eqref{eq_conn} show that the explicit simulation of the incoming light pulse circumvents the calculation of a four-point correlation function and reduces the problem to the measurement of a two-point correlation function.  

Alternatively, the formula for the resonant Raman scattering signal can be obtained by considering the bosonic Green's function of the outgoing photon, and its equation of motion. This has the advantage that it can be expressed using the L-shaped Keldysh contour for an initial state at temperature $T>0$.\cite{Aoki2014}
We define the photon Green's function
\begin{align}
B_\gamma(t,t')=&-i\langle \mathcal{T}_\mathcal{C}a_\gamma(t)a^\dagger_\gamma(t')\rangle,
\end{align}
from which the photon number (Raman signal) at time $t$ can be extracted as 
$\langle n_\gamma\rangle(t) =iB_\gamma^<(t,t)$. 
The photon Green's function satisfies the Dyson equation 
\begin{align}
(i\partial_t-\omega_\gamma)B_\gamma(t,t')-( g_\gamma^2\Pi_\text{res} \ast B_\gamma)(t,t')=\delta(t,t'), 
\end{align}
where the correlation function  $g_\gamma^2\Pi_\text{res}(t,t')$ now takes the role of the self energy to second oder in $g_\gamma$. (The linear order self-energy is a classical source field and is excluded for the same reason as the disconnected parts of the diagram in Eq.~\eqref{eq_raman}.) The number of emitted photons can thus be obtained from
\begin{align}
\frac{d}{dt}\langle n_\text{res}^{\gamma} \rangle(t)&= i\Big(\frac{d}{dt}+\frac{d}{dt'}\Big)B^<_\gamma(t,t')\Big|_{t=t'}\nonumber\\
&=g_\gamma^2(\Pi_\text{res}\ast B_\gamma-B_\gamma\ast\Pi_\text{res})^<(t,t). \label{dndt}
\end{align}
In the convolutions in Eq.~\eqref{dndt} we can then to leading order in $g_\gamma$ approximate $B_\gamma$ by the free photon Green's function $B_{0,\gamma}$. The convolutions on the contour $\mathcal{C}$ are evaluated using the standard Langreth rules,\cite{Stefanucci_book,Nessi} see Appendix~\ref{app_equivalence}. 

In practice, Eq.~\eqref{dndt} is more convenient for a numerical evaluation than Eq.~\eqref{eq_raman}, which requires the knowledge of the functions on the full contour, and thus even for times before the Raman pulse hits the system. The difference is that in the $T>0$ result \eqref{dndt} the outgoing photon Green's function has an initial state occupation given by the Bose function $n_B(\omega_{\rm out})$. However, this number is exponentially small in all relevant applications below. (Note that if the temperature were so high that the outgoing mode is initially occupied, this thermal occupation would also matter experimentally, and the rate should be computed with Eq.~\eqref{dndt} rather than Eq.~\eqref{eq_raman}.)   
In Appendix~\ref{app_equivalence}, we give explicit expressions for the contour convolutions on the $L$-shaped contour, and demonstrate that at $T=0$ Eqs.~\eqref{eq_raman} and \eqref{dndt} are related by an analytical continuation.

\subsection{Evaluation of $\Pi$ and difference between the four-point and two-point formalism}
\label{sec_4vs2point}

To calculate the photon emission rate using Eqs.~\eqref{dndt} and \eqref{eq_convolution}, we need to measure the correlation function \eqref{eq_conn}. Note that Eq.~\eqref{eq_conn} is, up to a prefactor $1/2$, the current-current correlation function. In equilibrium, within DMFT, the latter can be evaluated without vertex corrections,\cite{Kurana1990} 
which yields the bubble 
\begin{align}
\Pi_\text{res}(t,t') =  -i \sum_{k\sigma} M^{\gamma}_k(t) G_{k\sigma}(t,t') G_{k\sigma}(t',t) M^{\gamma}_k(t'),\label{eq_bubble}
\end{align}
where $G_{k\sigma}=-i\langle \mathcal{T}_\mathcal{C} c_{k\sigma}(t)c^\dagger_{k\sigma}(t')\rangle$ is the lattice Green's function of the system. The vanishing of vertex corrections relies on the fact that (i) the self-energy of the infinite-dimensional system is local, (ii) the current vertex $v_k$ is an odd function of momentum, and (iii) the Green's functions are even in momentum.\cite{Kurana1990,Aoki2014} The third point holds no longer in the presence of external fields, which is why vertex corrections can become important in the optical conductivity of driven systems.\cite{Eckstein2008,Tsuji2009} Since the correlation function \eqref{eq_conn} is evaluated in the presence of the external Raman probe, the factorization in Eq.~\eqref{eq_bubble} is an approximation (even in the limit $A_0\to 0$, as the full result is of order $A_0^2$). 

\begin{figure}[t]
\begin{center}
\includegraphics[angle=0, width=\columnwidth]{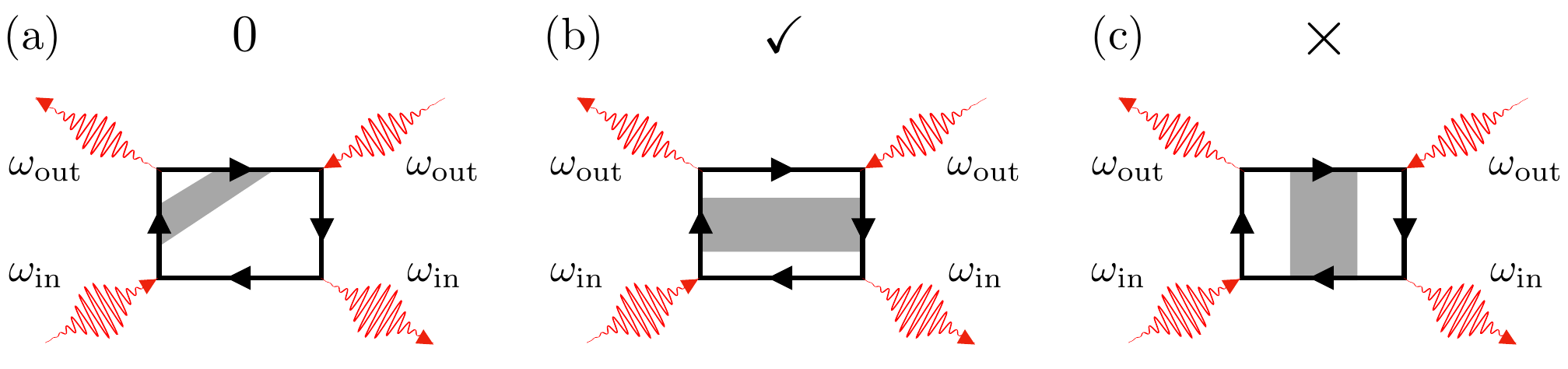}
\caption{
Illustration of different types of vertex corrections to the resonant Raman scattering process. Vertex corrections of the type shown in panel (a) vanish in the absence of pump light and for weak probe pulses, those illustrated in panel (b) are taken into account, while the vertex corrections shown in panel (c) are neglected. 
}
\label{fig_vertex}
\end{center}
\end{figure}   

It is therefore illustrating to discuss the type of vertex corrections which are included in the expression \eqref{eq_bubble} by considering the implications of this approximation for the resonant Raman diagram. To get from the two-time bubble to the conventional four-time resonant Raman diagram, one can, in the perturbative regime, explicitly represent the interaction with the incoming light in the propagators. This leads to diagrams as shown in Fig.~\ref{fig_vertex}, where the outgoing light marks the time arguments of the bubble. 
In the absence of pump light and for weak probe pulses, the vertex corrections shown in Fig.~\ref{fig_vertex}(a) vanish because of points (ii) and (iii) mentioned above. 
The vertex corrections illustrated in Fig.~\ref{fig_vertex}(b) are contained in the bubble calculation, since the Green's functions include the effect of the Raman drive, while vertex corrections of the type shown in panel (c) are neglected. A complete analysis of vertex corrections is beyond the scope of this paper, and will be deferred to future work. 
However, already the evaluation of the resonant Raman diagrams without full vertex corrections shows that these contributions can be of similar order as the nonresonant diagrams and should matter for the interpretation of experiments. 
The results below will thus be calculated with the bubble approximation \eqref{eq_bubble}.

Let us add a few comments on the benefits of the two-point formalism. In principle the four-point ring diagrams could be evaluated by first numerically calculating the dressed Green's function to leading order in $A_0$, via a convolution, and then applying the two-point formalism. Nevertheless, the two-point formalism is easier to implement and more convenient when additional vertex corrections are considered. In fact, it is relatively straight-forward to build in further vertex corrections into the two-point bubble, for example through ladder corrections which represent the interactions of electrons with spin fluctuations~\cite{Kauch2020,Simard2021,Worm2021,Simard2021b} or superconducting fluctuations.\cite{Aslamazov1968} 
Another advantage is that the expressions derived in this work can be extended directly to wave function methods.

While we focus in the main text on resonant Raman scattering, other contributions to the Raman signal can be derived in an analogous fashion. In Appendix~\ref{app_nonresonant} we derive the standard diagrammatic expression for nonresonant Raman scattering, and show some applications. 

\section{Results}
\label{sec:results}

To demonstrate our formalism for the simulation of resonant Raman scattering we consider the Holstein-Hubbard model (\ref{eq_H}) on an infinite-dimensional hypercubic lattice with a noninteracting density of states $\rho(\omega)=\exp(-\omega^2/W^2)/(\sqrt{\pi}W)$, and measure energy (time) in units of $W$ ($\hbar/W$). The polarization of the incoming field in the Raman measurement is chosen along the body diagonal, $\epsilon_\alpha=(1,1,1,\dots)$,  and the same holds for the pump laser field in the nonequilibrium simulations. In the following, the light-matter coupling parameter $\frac{g_\gamma}{\sqrt{2}}$, the electron charge $q$, the dielectric constants and $\hbar$ are all set to $1$. 

The hypercubic lattice has a bare dispersion 
\begin{equation}
\varepsilon_k = -2t^*\sum_{\nu=1}^d \cos(k_\nu)
\end{equation}
with $t^*=\frac{W}{2\sqrt{d}}$ and $d\rightarrow\infty$. The effect of the Peierls phase is to shift the dispersion as
\begin{equation}
\varepsilon_{k-A(t)}=\cos(A(t))\varepsilon_k+\sin(A(t))\bar \varepsilon_k.
\end{equation}
with $\bar \varepsilon_k=-2t^*\sum_{\nu=1}^d\sin(k_\nu)$. Momentum summations can thus be implemented as integrals over the joint density of states $D(\varepsilon,\bar\varepsilon)=\sum_{k}\delta(\varepsilon-\varepsilon_k)\delta(\bar\varepsilon-\bar\varepsilon_k)$, which in the case of the hypercubic lattice factorizes as $D(\varepsilon,\bar\varepsilon)=D(\varepsilon)D(\bar \varepsilon)$.\cite{Turkowski2005} 

We solve the Holstein-Hubbard model on the hypercubic lattice using DMFT and the noncrossing approximation (NCA) as the impurity solver.\cite{Eckstein2010,Werner2013} 
The NCA is meaningful in the strong electron-phonon coupling (polaronic) regime. 
Our nonequilibrium DMFT setup is the same as the one used in Ref.~\onlinecite{Werner2015}.

The choice of the polarization vectors $\epsilon$ determines the type of Raman response. 
The $A_{1g}$ channel corresponds to $\epsilon_\alpha=\epsilon_{\gamma}=(1,1,1,\ldots)$ and the $B_{1g}$ channel to $\epsilon_\alpha=(1,1,1,\ldots)$ and $\epsilon_{\gamma}=(-1,1,-1,1,\dots)$, so that\cite{Shvaika2005} 
\begin{align}
M^{A_{1g}}_k&\propto \frac{1}{\sqrt{d}}\sum_{\mu}\sin({k_\mu}),\\
M^{B_{1g}}_k&\propto \frac{1}{\sqrt{d}}\sum_{\mu}(-1)^\mu\sin({k_\mu}).
\end{align}
In the $d\rightarrow \infty$ limit, $M^{B_{1g}}_k$ can be replaced by a constant, as discussed in Appendix~\ref{app_B1g}. The simulation results in this paper are for the $A_{1g}$ response.

\subsection{Equilibrium results}

\begin{figure}[t]
\begin{center}
\includegraphics[angle=-90, width=0.8\columnwidth]{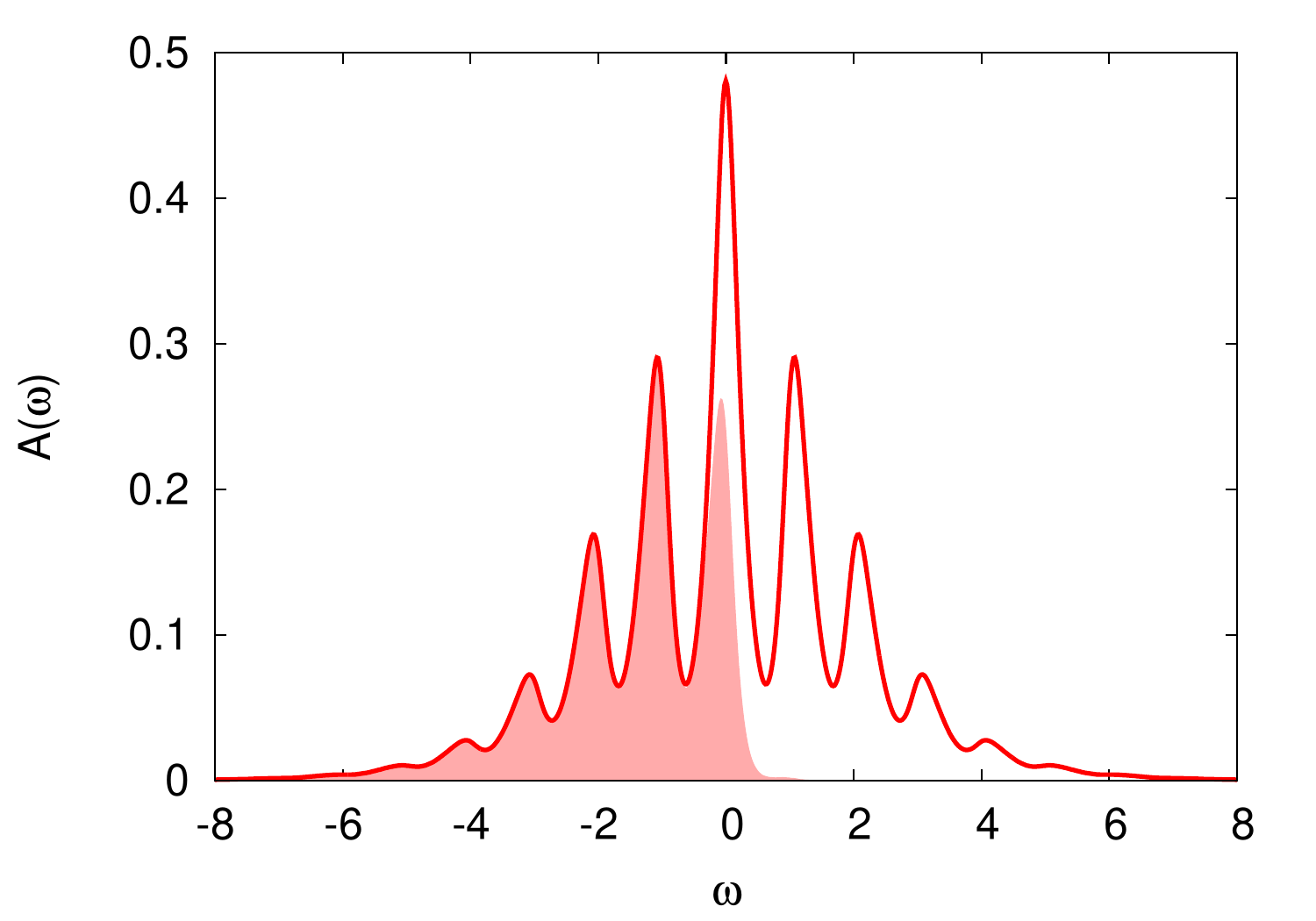}
\caption{DMFT+NCA spectral function of the Holstein-Hubbard model with $U=2$, $g=1$, $\omega_0=1$ and inverse temperature $\beta=5$. The occupied part of the spectrum is indicated by the shading. 
}
\label{fig_eq_u2}
\end{center}
\end{figure}

To illustrate our measurement procedure, we first consider the equilibrium Holstein-Hubbard model in the metallic phase. If the electron-phonon coupling is strong enough, the local spectral function features peaks corresponding to polaron subbands, as illustrated for $U=2$, $g=1$, $\omega_0=1$, and inverse temperature $\beta=5$ in Fig.~\ref{fig_eq_u2}. The main peak at $\omega=0$ corresponds to electron insertion or removal processes that do not excite phonons, while those at energies $\omega\approx n\omega_0$ ($n\ne 0$) involve the additional excitation of phonons. The solid line plots the spectral function $A(\omega)=-\frac{1}{\pi}\text{Im}G^R(\omega)$, while the shading indicates the occupied part of the spectral function. In equilibrium, the occupation is given by $A^<(\omega)=A(\omega) n_F^\beta(\omega)$, with $n_F^\beta(\omega)$ the Fermi function for inverse temperature $\beta$. 

\begin{figure}[t]
\begin{center}
\includegraphics[angle=-90, width=0.8\columnwidth]{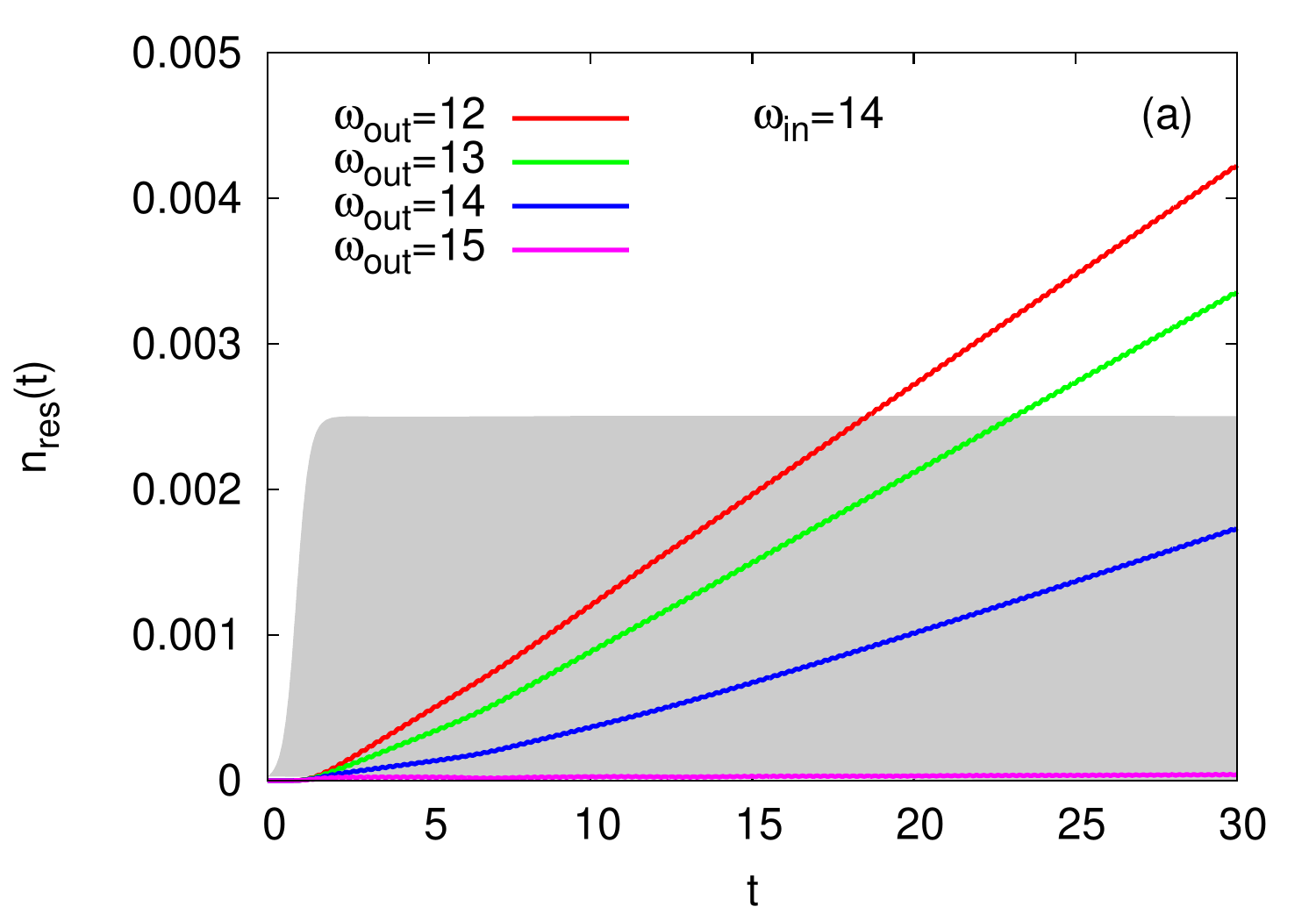}
\includegraphics[angle=-90, width=0.8\columnwidth]{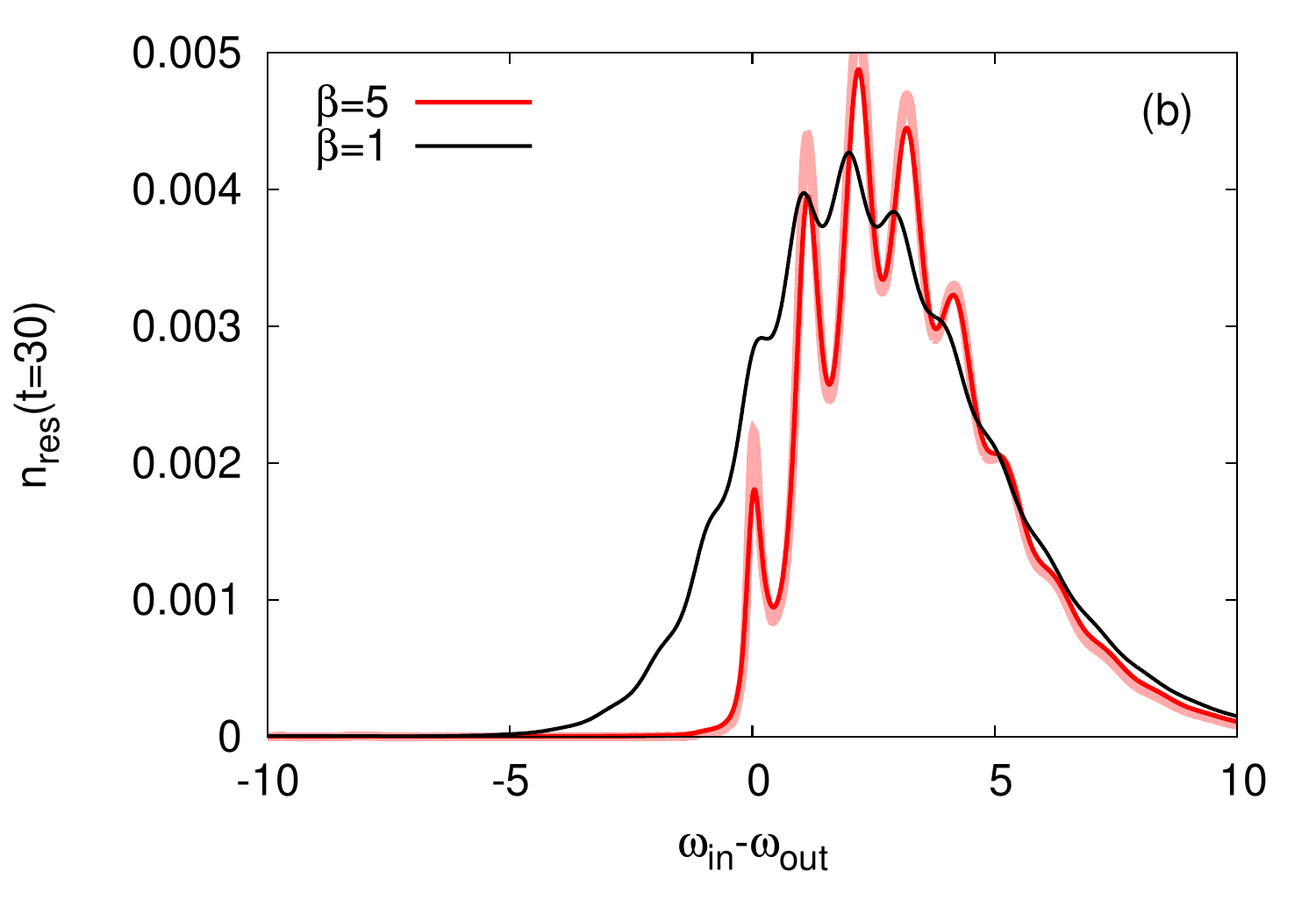}
\caption{Resonant contribution to the Raman signal for $U=2$, $g=1$, $\omega_0=1$ and inverse temperature $\beta=5$ (photon number measured at time $t$). The Raman drive has frequency $\omega_\text{in}=14$ and amplitude $\mathcal{E}_\text{in}=0.5$. Panel (a) shows the time evolution of the number of emitted photons for the indicated values of $\omega_\text{out}$, together with a rescaled envelope of the Raman drive (gray shading). Panel (b) shows the corresponding spectrum (proportional to the resonant Raman signal) measured at time $t=30$ for $\beta=5$ (red line) and $\beta=1$ (black line). The thick light-red line is the spectrum for $\beta=5$ obtained by averaging the rate \eqref{dndt} over the time interval $27\le t \le 30$ (rescaled by a factor $29$ to match the scales).
}
\label{fig_eq_nraman_u2}
\end{center}
\end{figure}

In Fig.~\ref{fig_eq_nraman_u2} we plot the evolution of $n_\text{res}(t)$, evaluated with Eqs.~\eqref{dndt} and \eqref{eq_convolution}, for different $\omega_\text{out}$. Here, the Raman drive 
\begin{equation}
E_\text{in}(t)=\mathcal{E}_\text{in} f_\text{in}(t)\sin(\omega_\text{in}t) \label{E_Raman}
\end{equation}
is simulated by a periodic electric field with $\omega_\text{in}=14$ (near-resonant regime) and amplitude $\mathcal{E}_\text{in}=0.5$, which is ramped on in a time $\lesssim 2$ and continues up to the longest simulation time $t=30$. The form of the envelope function $f_\text{in}(t)$ is indicated by the gray shading in panel (a).  

\begin{figure}[t]
\begin{center}
\includegraphics[angle=-90, width=0.8\columnwidth]{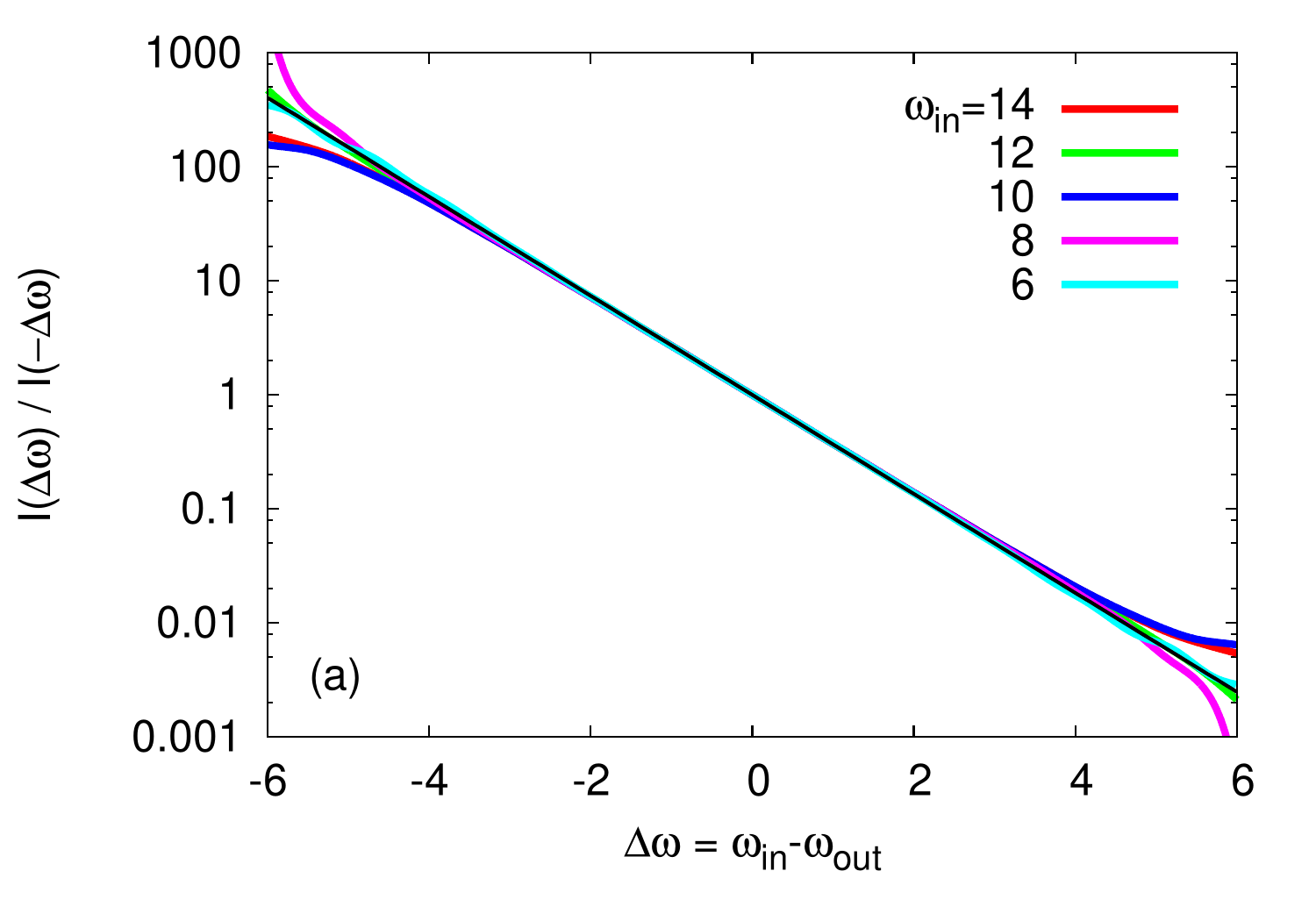}
\includegraphics[angle=-90, width=0.8\columnwidth]{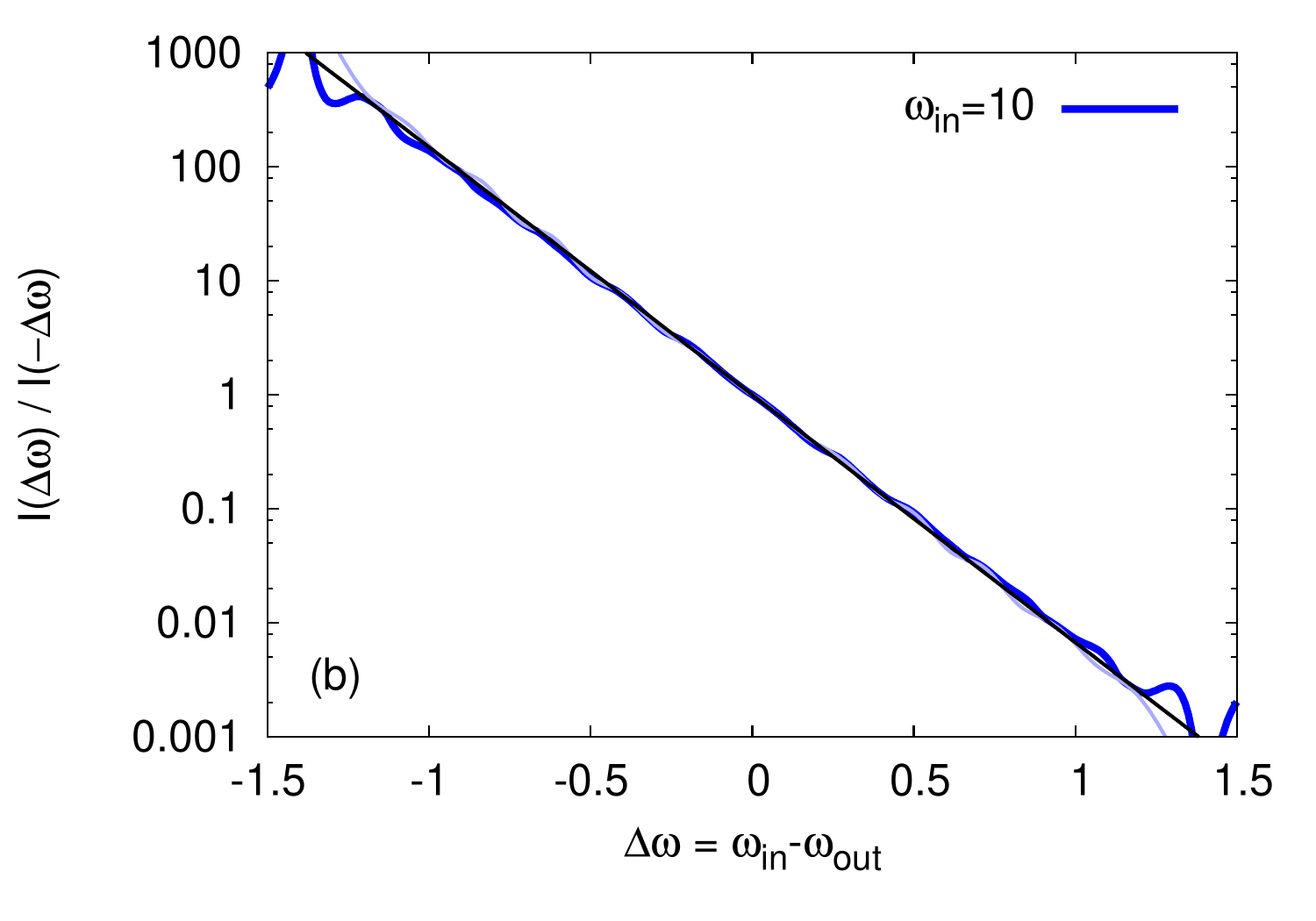}
\caption{
Ratio between the anti-Stokes signal $I_a=I(-\Delta\omega)$ and Stokes signal $I_s=I(\Delta\omega)$ (same setup as Fig.~\ref{fig_eq_nraman_u2}) as a function of $\Delta\omega=\omega_\text{in}-\omega_\text{out}$ on a logarithmic scale, and comparison to $e^{-\beta\Delta\omega}$ (black line). We consider the resonant contribution to the Raman signal for $U=2$, $g=1$, $\omega_0=1$ and inverse temperature $\beta=1$ (panel (a)) and $\beta=5$ (panel (b)). The results plotted are for driving amplitude $\mathcal{E}_\text{in}=0.5$. Panel (a) is based on the photon numbers measured at $t=30$ and panel (b) on the rates averaged over $27\le t\le 30$ (thick line) and $28.5\le t \le 30$ (thin line). 
For $\beta|\Delta \omega| \gtrsim 4$ the numerical uncertainty on the $I_a/I_s$ ratio becomes large. 
}
\label{fig_eq_nraman_u2_ratio}
\end{center}
\end{figure}

As can be seen in Fig.~\ref{fig_eq_nraman_u2}(a), the number of emitted photons grows approximately linearly with time, with a slope that depends on $\omega_\text{out}$. For $\omega_\text{in}-\omega_\text{out}<0$ (anti-Stokes processes), the Raman signal is small, while for $0<\omega_\text{in}-\omega_\text{out}\lesssim \text{bandwidth}$ (Stokes processes) the signal is larger. A spectrum proportional to the resonant Raman signal can be obtained by plotting $n_\text{res}(t)$ for fixed time $t$ as a function of $\omega_\text{in}-\omega_\text{out}$. The result for $\beta=5$ and $t=30$ is shown by the red line in Fig.~\ref{fig_eq_nraman_u2}(b). For comparison, we also plot by a black line the analogous result obtained for $\beta=1$. At this higher temperature, the anti-Stokes peaks become more prominent. 

In Fig.~\ref{fig_eq_nraman_u2}(b) we also indicate by a thick light-red line the result obtained by averaging the rate \eqref{dndt} over a small time interval ($27\le t \le 30$). After a rescaling, the result looks consistent with the spectrum measured at $t=30$, but the substructures related to the phonon sidebands are sharper. Also on the anti-Stokes side, the exponentially suppressed signal is better captured by the averaged rate, while the photon number at fixed $t$ has a significant uncertainty due to small oscillations. 

Raman scattering is sometimes used for thermometry applications. In the case of a {\it nonresonant} signal, in equilibrium, the values of the Stokes (anti-Stokes) signals $I_s$ ($I_a$) at energy $\Delta\omega=\omega_\text{in}-\omega_\text{out}$ ($-\Delta\omega$) allow to extract the inverse temperature $\beta$ via the formula\cite{Devereaux2007} 
\begin{equation}
\frac{I_a}{I_s}=e^{-\beta\Delta\omega}.\label{eq_thermo}
\end{equation}
In the case of a {\it resonant} light-scattering process, the validity of Eq.~\eqref{eq_thermo} is a priori not clear. We test the relation in Fig.~\ref{fig_eq_nraman_u2_ratio}, which plots $I_a/I_s$ as a function of $\Delta\omega$ on a logarithmic scale and compares the result to $e^{-\beta\Delta\omega}$ (black line). Panel (a) shows results for $\beta=1$ and different $\omega_\text{in}$ in the near-resonant to resonant regime. Here, we show results based on the phonon numbers measured at $t=30$, since the accuracy of this measurement is sufficient at $\beta=1$. A reasonable numerical estimation of $I_a/I_s$ is possible in the energy range $-4\lesssim \Delta\omega\lesssim 4$, and we see that in this range, the results are consistent with Eq.~\eqref{eq_thermo}, even for the smallest $\omega_\text{in}$. Panel (b) shows results for $\beta=5$ and $\omega_\text{in}=10$. Here, because of the exponentially suppressed anti-Stokes signal, we use the averaged rate to estimate the ratio $I_a/I_s$. The two blue lines in the figure show the results obtained using different time intervals for the averaging. The deviations between them give an indication of the accuracy of the numerical estimate. Within this accuracy, Eq.~\eqref{eq_thermo} is again well satisfied. Similar results are found also for the larger and smaller $\omega_\text{in}$. At least for the Holstein-Hubbard model, with equal spacing $\omega_0$ between the phonon side-bands, we thus find that a determination of the temperature via Eq.~\eqref{eq_thermo} is still possible even if resonant scattering processes play an important role.

\begin{figure}[t]
\begin{center}
\includegraphics[angle=-90, width=0.8\columnwidth]{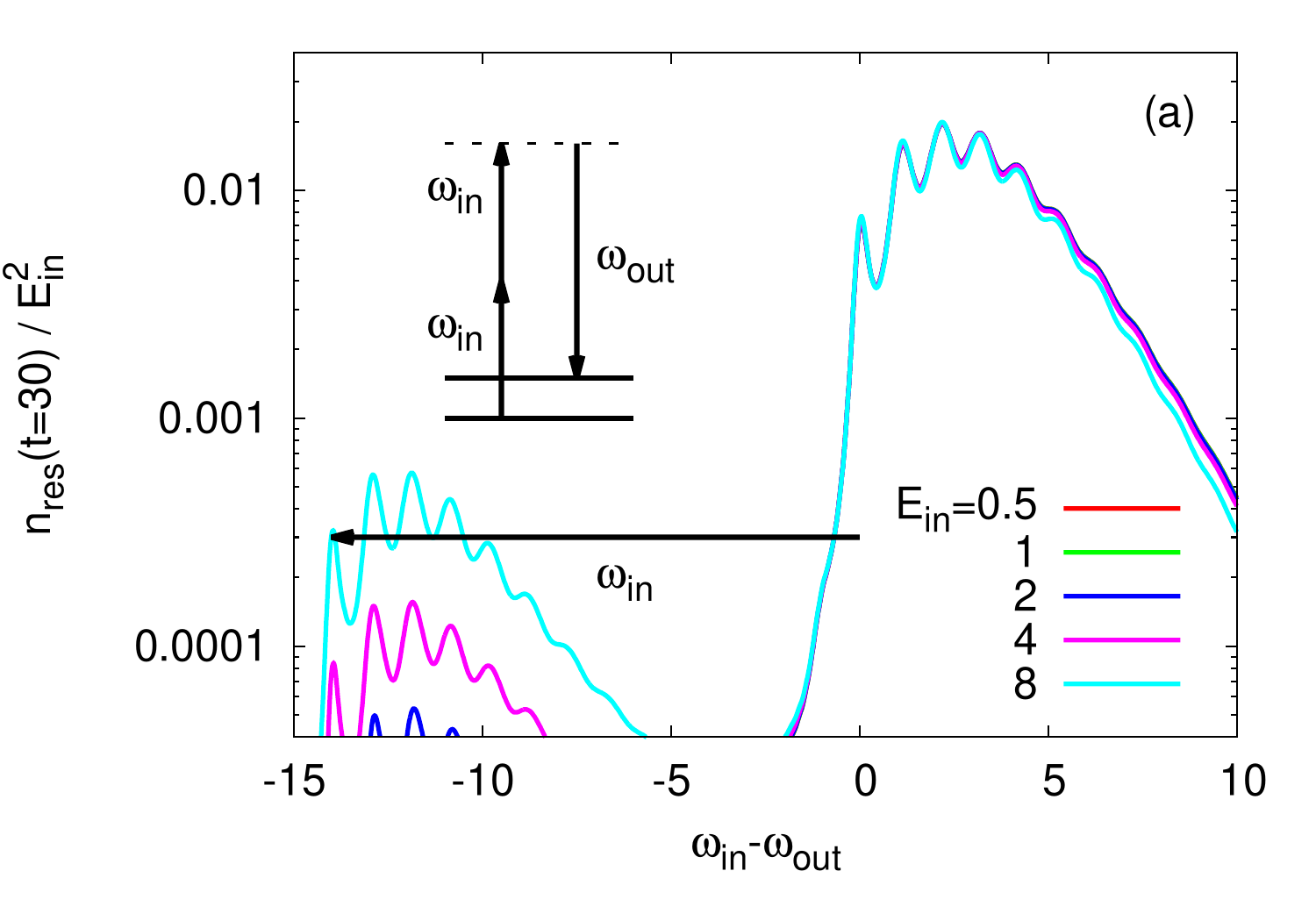}
\includegraphics[angle=-90, width=0.8\columnwidth]{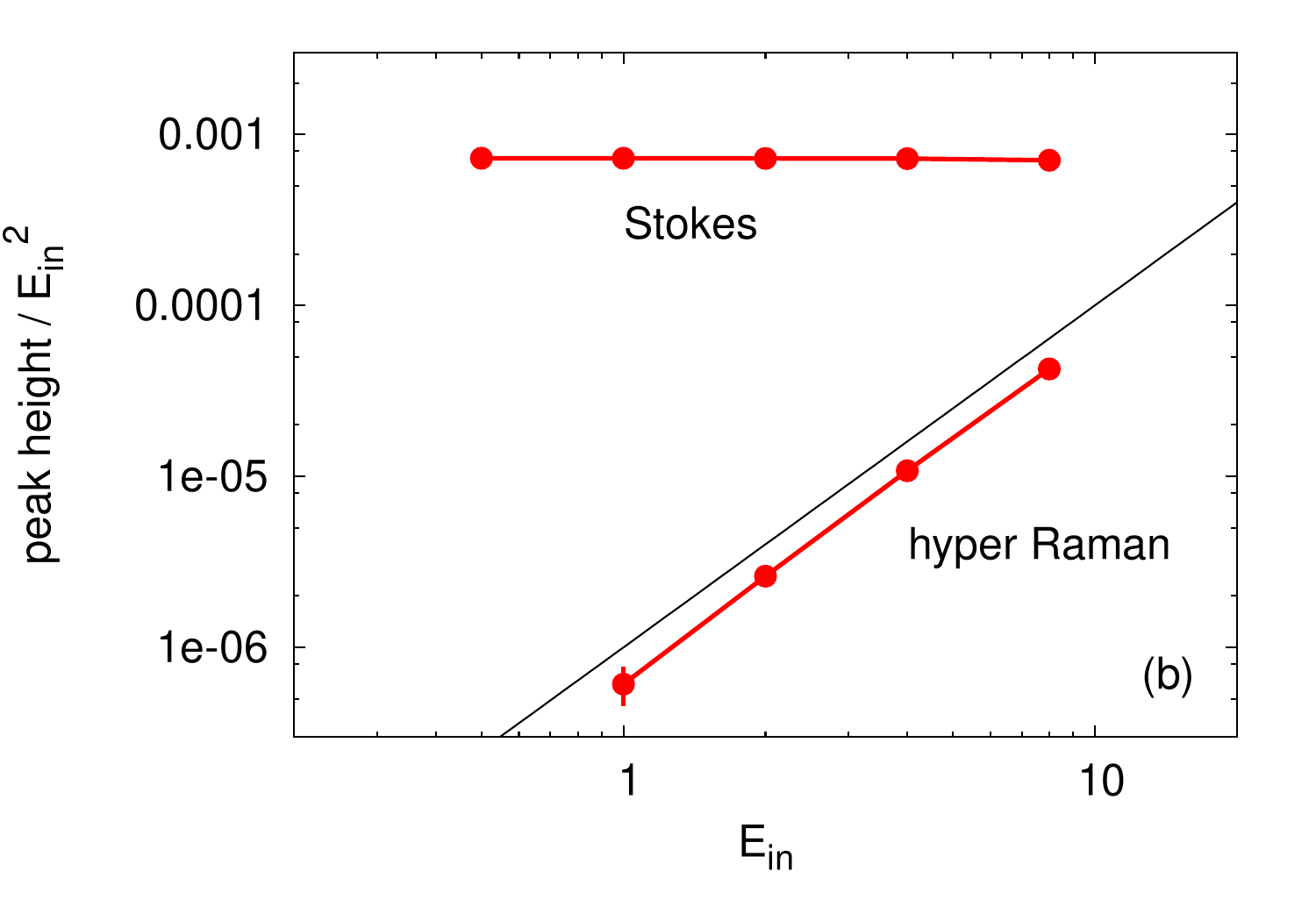}
\caption{
Panel (a):
Resonant contribution to the Raman signal (rescaled by $\mathcal{E}_\text{in}^{-2}$) for $U=2$, $g=1$, $\omega_0=1$ and inverse temperature $\beta=5$. The Raman drive has frequency $\omega_\text{in}=14$, while the amplitude is varied from $\mathcal{E}_\text{in}=0.5$ to $8$. For strong probe field amplitudes, Stokes peaks associated with two-photon absorption appear on the $\omega_\text{in}-\omega_\text{out}<0$ side (hyper Raman signal, see inset). Panel (b): Scaling of the highest peaks in the Stokes and hyper Raman signal, rescaled by $\mathcal{E}_\text{in}^{-2}$, as a function of $\mathcal{E}_\text{in}$. The black line is proportional to $\mathcal{E}_\text{in}^2$. The data in (b) have been extracted from the rate \eqref{dndt} averaged over $27\le t \le 30$.
}
\label{fig_amplitude}
\end{center}
\end{figure}

The effect of the amplitude $\mathcal{E}_\text{in}$ of the incoming field is illustrated in Fig.~\ref{fig_amplitude}. Panel (a) plots results analogous to the red line in Fig.~\ref{fig_eq_nraman_u2}(b), but now for different probe field amplitudes $\mathcal{E}_\text{in}$, and with the signal rescaled by $\frac{1}{\mathcal{E}_\text{in}^2}$. In the weak-field regime, these rescaled spectra overlap, confirming the expected quadratic scaling in the field (Eq.~\eqref{Gamma_out}). In particular, the field strength $\mathcal{E}_\text{in}=0.5$ used in most of our calculations is, as far as the Stokes signal is concerned, well within the perturbative regime. On the anti-Stokes side $\omega_\text{in}-\omega_\text{out}<0$ one observes a different scaling. At stronger probe-field amplitudes, a prominent series of peaks appears down to an energy of approximately $-\omega_\text{in}$. These are processes associated with second-order photon absorption from the Raman drive, as illustrated in the inset in Fig.~\ref{fig_amplitude}(a). In the literature, these peaks are also referred to as hyper Raman signal.\cite{Madzharova2017} In panel (b), we analyze the scaling with $\mathcal{E}_\text{in}$ of the highest peak for positive and negative $\omega_\text{in}-\omega_\text{out}$. Because of the very weak hyper Raman signal for small $\mathcal{E}_\text{in}$, we base this analysis on the rates averaged in the interval $27\le t\le 30$. 
While for the Stokes signal and ``regular" anti-Stokes signal, the peak height times $\mathcal{E}_\text{in}^{-2}$ is approximately constant in the considered field range, near $\omega_\text{in}-\omega_\text{out} \approx -\omega_\text{in}$, we observe an approximately quadratic increase, which means that the total scaling of this hyper Raman feature is $\propto\mathcal{E}_\text{in}^4$, 
which is consistent with the second-order absorption. 

\begin{figure}[t]
\begin{center}
\includegraphics[angle=-90, width=0.8\columnwidth]{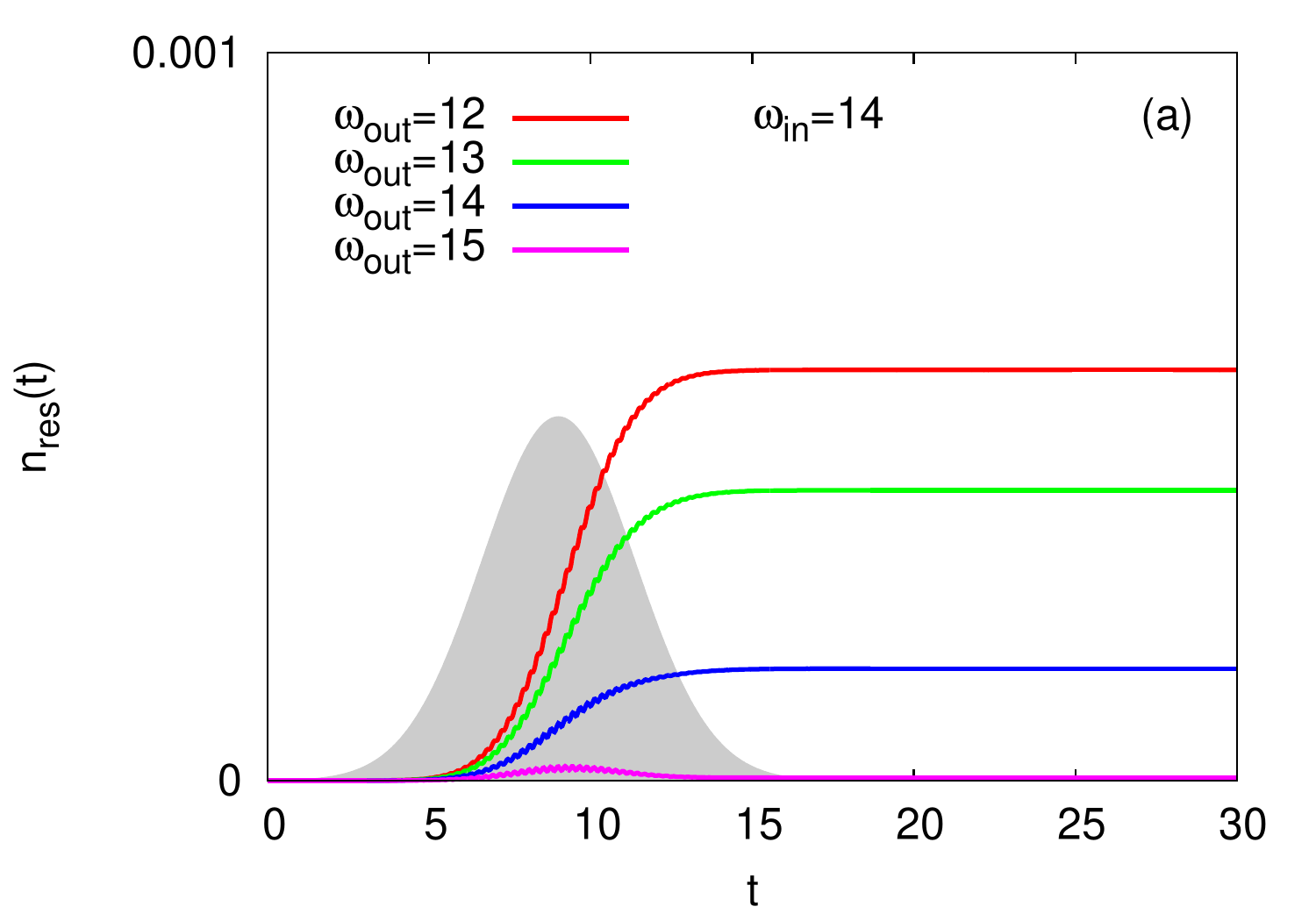}
\includegraphics[angle=-90, width=0.8\columnwidth]{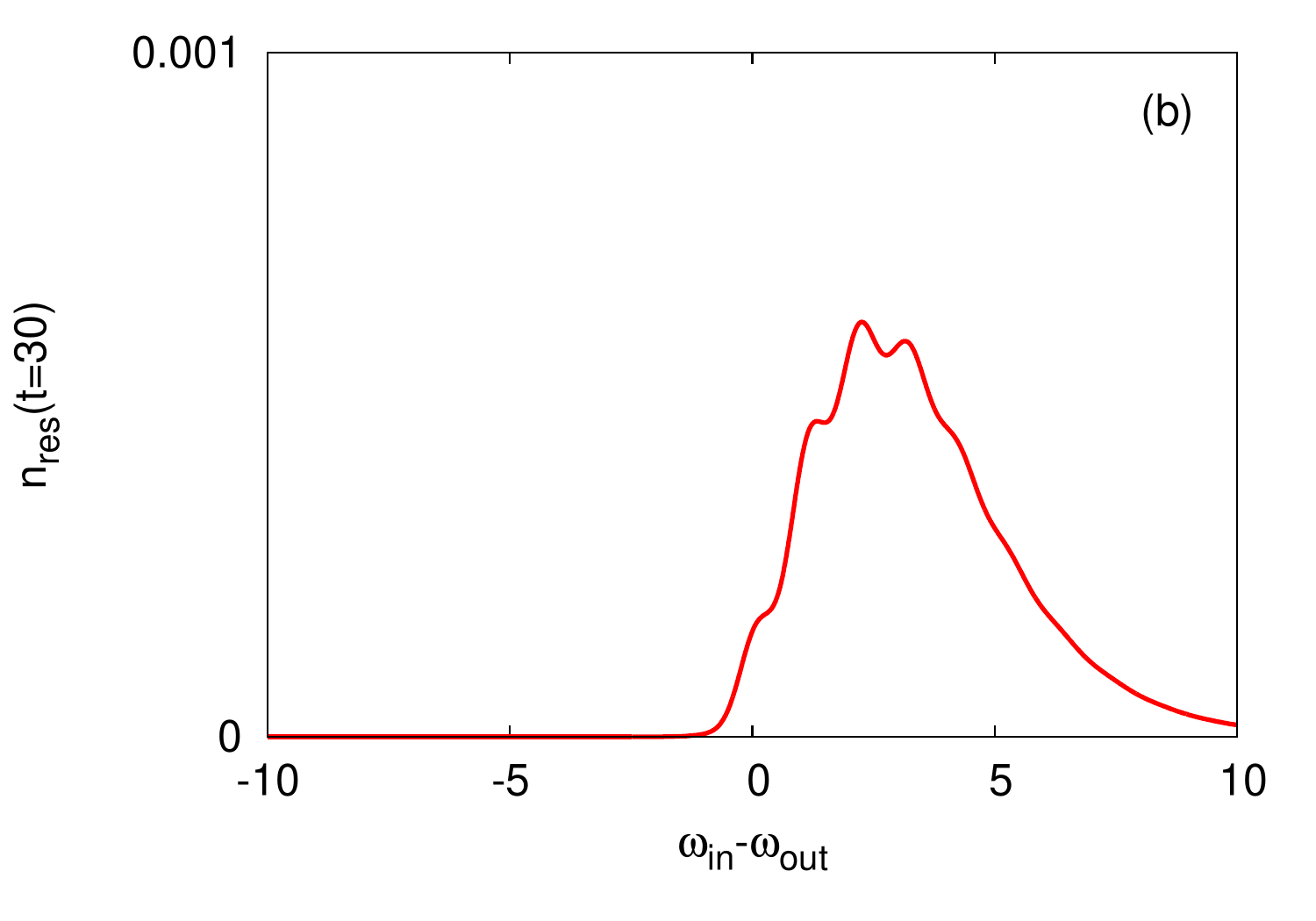}
\caption{Resonant contribution to the Raman signal for $U=2$, $g=1$, $\omega_0=1$ and inverse temperature $\beta=5$ (photon number measured at time $t$), for a short Raman probe pulse. The Raman drive has frequency $\omega_\text{in}=14$ and amplitude $\mathcal{E}_\text{in}=0.5$. Its (rescaled) envelope function is indicated by the gray shading in panel (a). The lines in panel (a) show the time evolution of the number of emitted photons as a function of time for the indicated values of $\omega_\text{out}$, while panel (b) shows the Raman spectrum measured at time $t=30$.   
}
\label{fig_eq_nraman_u2_pulse}
\end{center}
\end{figure}

In Fig.~\ref{fig_eq_nraman_u2_pulse} we show results analogous to Fig.~\ref{fig_eq_nraman_u2} (for $\beta=5$), but now obtained using a short Raman probe pulse with a Gaussian envelope $f_\text{in}(t)=\exp(-0.3 (t-t_\text{probe})^2)$ centered at $t_\text{probe}=9$. A rescaled version of this envelope is indicated by the gray shading in panel (a). Outgoing photons are generated during the application of the probe pulse, while the signal $n_\text{res}(t)$ saturates after the probe pulse, as expected. The corresponding spectrum measured at $t=30$ is shown in panel (b). Because of the short probe pulse, the energy resolution is reduced, compared to Fig.~\ref{fig_eq_nraman_u2}, but one can still resolve peaks corresponding to different phonon sidebands. In the following section, we will employ such short probe pulses to study the nonequilibrium evolution of photo-excited systems. 

\begin{figure}[t]
\begin{center}
\includegraphics[angle=-90, width=0.8\columnwidth]{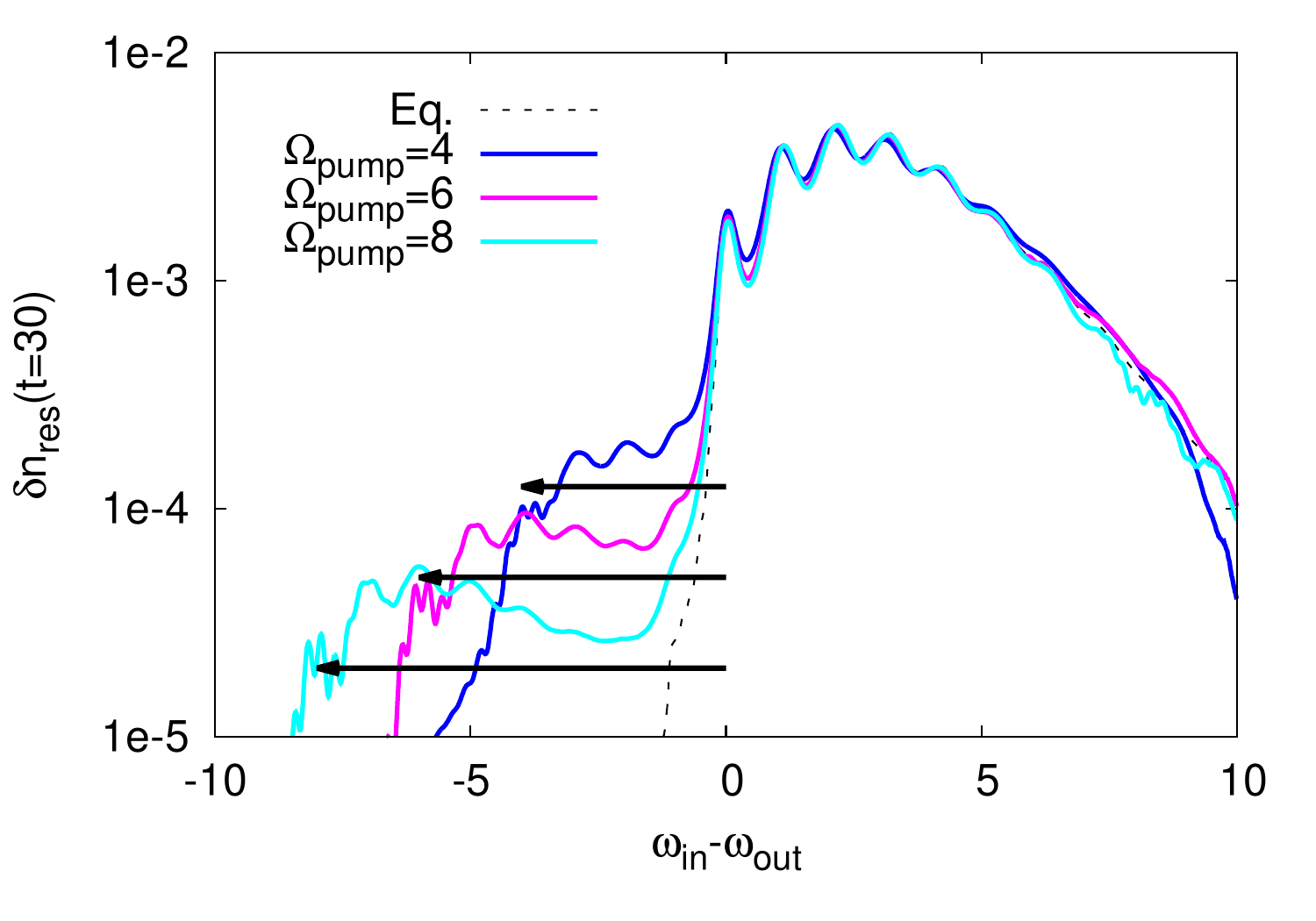}
\caption{Difference spectra $\delta n_\text{res}=\langle n_\text{res}\rangle_\text{pump+probe}-\langle n_\text{res}\rangle_\text{pump}$ for the pump-excited metallic system with $U=2$, $g=1$, and initial inverse temperature $\beta=5$. The photo-doping pulse has frequency $\Omega_\text{pump}=4,6,8$ and amplitude $\mathcal{E}_\text{pump}=4$. The Raman probe pulse has amplitude $\mathcal{E}_\text{in}=0.5$ and frequency $\omega_\text{in}=14$ (same as in Fig.~\ref{fig_eq_nraman_u2}). Black arrows indicate the energy range $-\Omega_\text{pump}\lesssim \omega_\text{in}-\omega_\text{out} \lesssim 0$. 
}
\label{fig_u2_wpump}
\end{center}
\end{figure}

\subsection{Nonequilibrium results}

\subsubsection{Photo-excited metallic system}
\label{sec_photo_metallic}

An advantage of the measurement approach based on real-time simulations is that it can easily deal with nonequilibrium states. In this section, we demonstrate this with results for the photo-excited Holstein-Hubbard model. In a metallic system, photo-excited electrons quickly relax due to scattering and the interactions with phonons,\cite{Sentef2013,Murakami2015} which results in a state which is close to an equilibrium state at higher $T$. During and shortly after the pump pulse, the electron and phonon populations are however highly nonthermal, and an interesting question is how this affects the Raman signal. 

We apply a pump pulse corresponding to the electric field
\begin{equation}
E_\text{pump}(t)=\mathcal{E}_\text{pump} f_\text{pump}(t-t_\text{pump})\sin(\Omega_\text{pump}(t-t_\text{pump}))
\end{equation}
with $\mathcal{E}_\text{pump}$ the peak amplitude, $\Omega_\text{pump}$ the frequency and $f_\text{pump}(t)$ the envelope of the pump pulse. To this strong perturbation we add in the simulation the weak Raman probe field \eqref{E_Raman} and calculate the difference 
\begin{equation}
\delta n_\text{res}=\langle n_\text{res}\rangle_\text{pump+probe}-\langle n_\text{res}\rangle_\text{pump}
\end{equation} 
of the photon numbers with pump and probe field, and with pump field only. 

Figure~\ref{fig_u2_wpump} shows the Raman spectra measured at $t=30$ in a system with $U=2$, $g=1$, $\omega_0=1$ and initial $\beta=5$, which is photoexcited by a pump pulse with pump frequency $\Omega_\text{pump}=4,6,8$ and peak amplitude $\mathcal{E}_\text{pump}=4$. The pump pulse is centered at time $t_\text{pump}=6$ and has a Gaussian envelope of the form $f_\text{pump}(t)=\exp(-0.3 t^2)$. The probe pulse is the same as in Fig.~\ref{fig_eq_nraman_u2}, i.e., the Raman drive continues with amplitude $\mathcal{E}_\text{in}=0.5$ up to the longest simulation time. We see that in contrast to the equilibrium system, whose Raman signal rapidly decays on the anti-Stokes side $\omega_\text{in}-\omega_\text{out}<0$ (black dashed line), the spectrum of the photo-doped system exhibits a plateau structure down to $\omega_\text{in}-\omega_\text{out}\approx -\Omega_\text{pump}$, as indicated by the arrows. This plateau could have two possible origins: (i) frequency mixing between the pump pulse and the Raman probe, or (ii) light scattering with energy gain from photo-excited states with excess kinetic and/or phonon energy up to $\Omega_\text{pump}$. The process (i) is analogous to hyper Raman scattering, but with absorption of $\Omega_\text{pump}+\omega_\text{in}$ instead of $\omega_\text{in}+\omega_\text{in}$. Alternatively, it can be viewed as Raman scattering from photon-dressed states (Floquet states).   
The fact that the shape of the ``plateau" resembles the Stokes signal for large $\Omega_\text{pump}$ suggests that at least in this limit, the frequency mixing dominates.
  
\begin{figure}[t]
\begin{center}
\includegraphics[angle=-90, width=0.8\columnwidth]{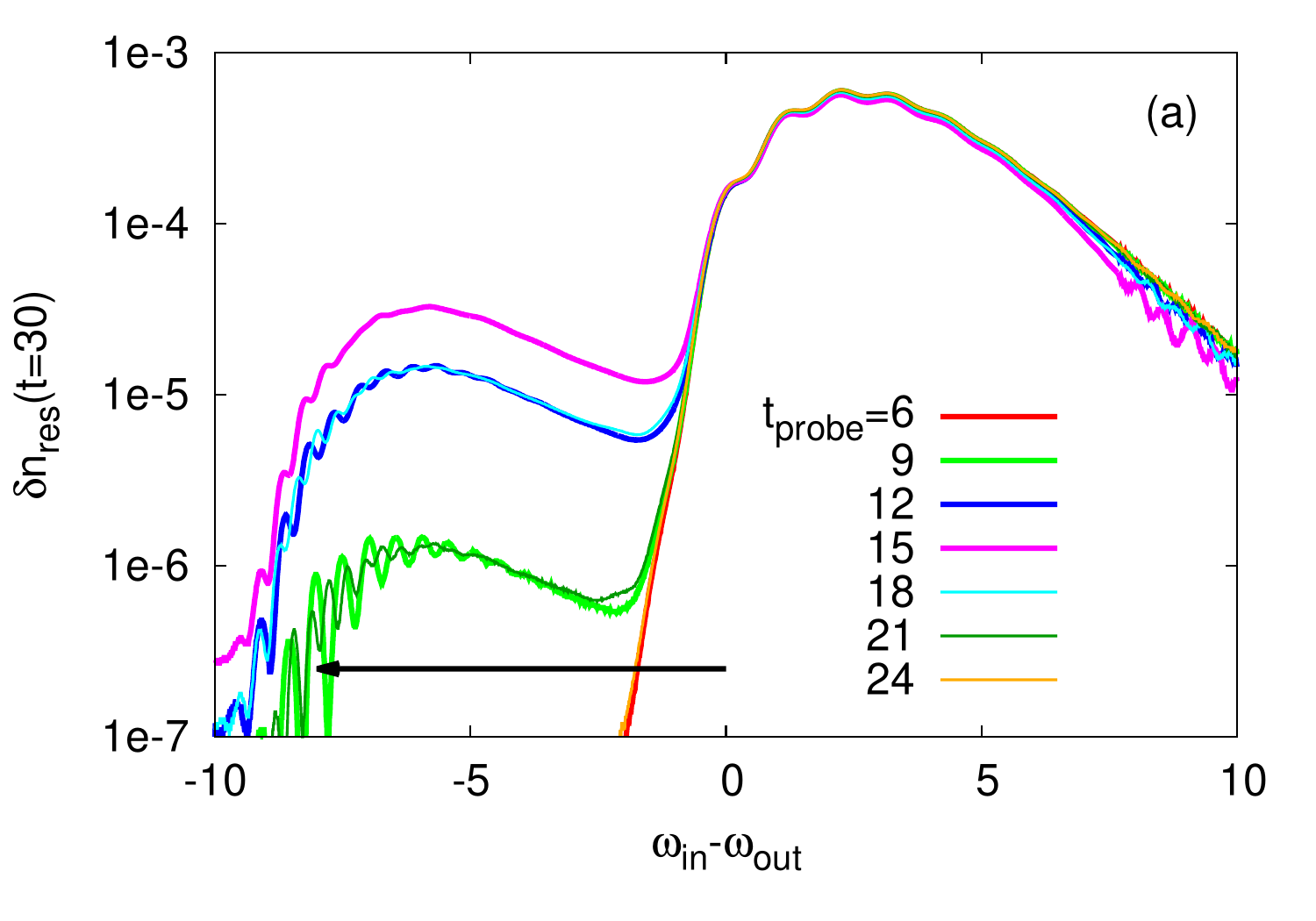}
\includegraphics[angle=-90, width=0.8\columnwidth]{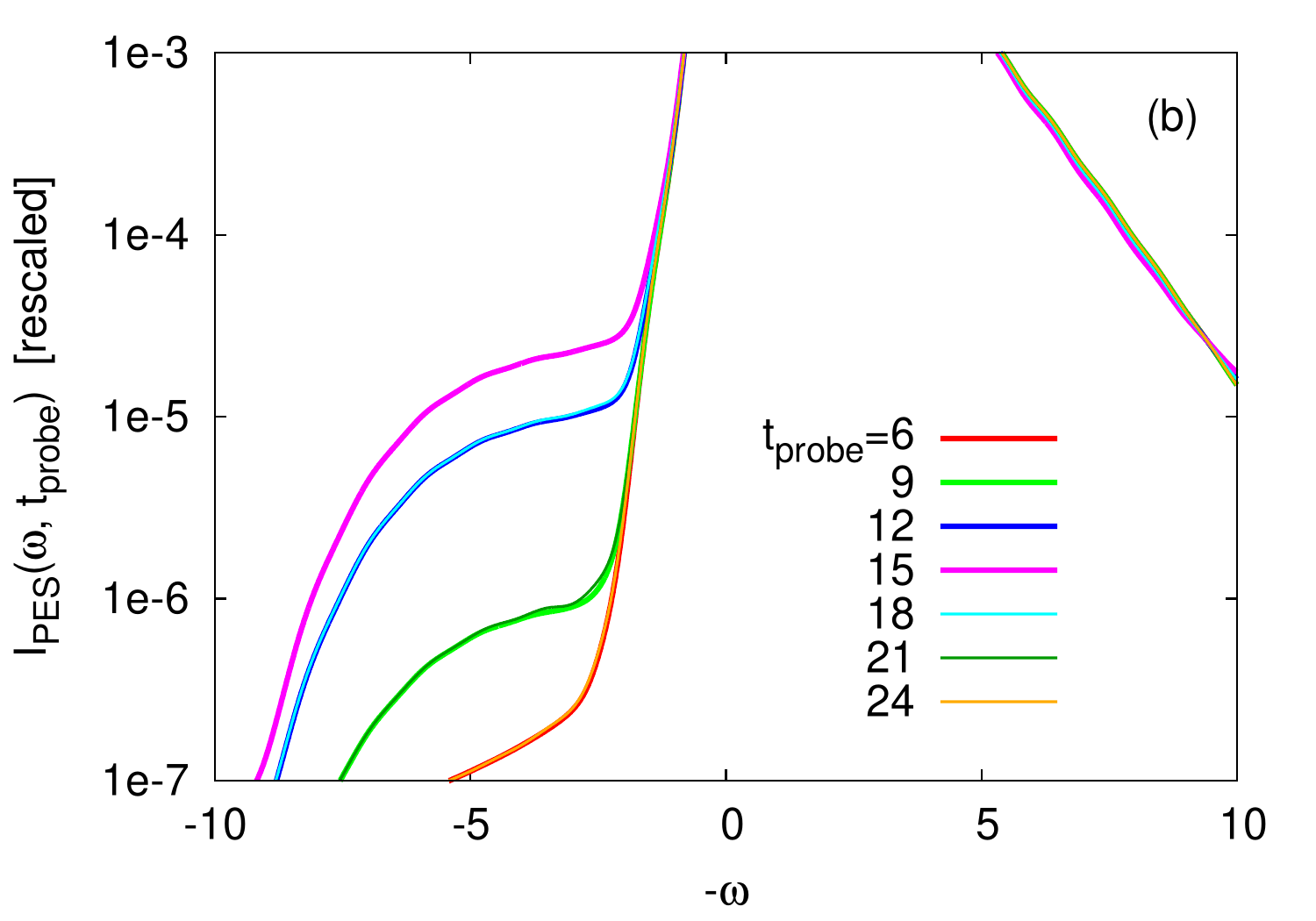}
\caption{Panel (a): Difference spectra $\delta n_\text{red}=\langle n_\text{res}\rangle_\text{pump+probe}-\langle n_\text{res}\rangle_\text{pump}$ for the pump excited metallic system with $U=2$, $g=1$, $\omega_0=1$ and initial $\beta=5$. The pump pulse has amplitude $\mathcal{E}_\text{pump}=4$, frequency $\Omega_\text{pump}=8$ and is centered at $t_\text{pump}=15$. The different lines show the resonant Raman signals obtained for the probe times $t_\text{pr}=6$, 9, \ldots, $24$. Both the pump and the probe pulses have a Gaussian envelope with the same form as shown by the gray shading in Fig.~\ref{fig_eq_nraman_u2_pulse}. The probe pulse has amplitude $\mathcal{E}_\text{in}=0.5$ and frequency $\omega_\text{in}=14$. The black arrow indicates the energy range $-\Omega_\text{pump}\lesssim \omega_\text{in}-\omega_\text{out} \lesssim 0$.
Panel (b): Time-resolved photoemission spectrum $I_\text{PES}(\omega,t_\text{probe})$ as a function of $-\omega$, for the same photodoped state and the same probe envelopes. 
}
\label{fig_u2_wpump_t}
\end{center}
\end{figure}

More insights into the population dynamics and the role of frequency mixing can be obtained by time-resolved measurements with short probe pulses. The results for probe pulses with a width $\Delta t\approx 6$ and a pump pulse with $E_\text{pump}$ centered at $t_\text{pump}=15$ are shown in Fig.~\ref{fig_u2_wpump_t}(a). The envelope of the pump pulse and the other system parameters are the same as before ($U=2$, $g=1$, $\beta=5$). The figure again plots the difference spectra $\delta n_\text{res}$, for the indicated probe times, which are symmetric with respect to $t_\text{pump}$. Because of the short probe pulses, the phonon peaks in the Raman spectra are washed out, but the measurements provide enough time resolution to track the growth and vanishing of the nonthermal plateau on the $\omega_\text{in}-\omega_\text{out}<0$ side of the spectrum. In particular, one can see that the results are symmetric with respect to $t_\text{pump}=15$. For example, the curves for $t_\text{probe}=12, 18$ or $t_\text{probe}=9, 21$ are almost identical. In the metallic phase of the Hubbard-Holstein model with strong electron-phonon coupling, one can expect a quick relaxation of the nonequilibrium photo-excited distributions to a thermal state with a temperature close to the initial temperature, which would be consistent with the quick return of the Raman signal to the initial state. However, the almost perfect temporal symmetry of the Raman signal with respect to the pump maximum, together with the spectral shape of the signal, suggest that there is in fact little contribution from photo-excited states, and that the Raman signal is instead predominantly due to photon-dressed states (frequency mixing).

The wiggles in the signal are due to the oscillations in $n_\text{res}(t)$ as a function of $t$. To suppress these, we average the signals over the time interval $27\le t \le 30$ in the case of $t_\text{probe}\le 21$. The anti-Stokes signal near $\omega_\text{in}-\omega_\text{out}\lesssim 0$ is however a bit broader after the pump maximum, compared to the corresponding signal before the pump maximum, which indicates some heating effect. The heating is small, because the phonon subsystem can absorb a large amount of energy.  

For a more direct view on the electronic population dynamics, 
we plot in Fig.~\ref{fig_u2_wpump_t}(b) the  time-resolved photoemission signal\cite{Freericks2009} 
\begin{align}
&I_\text{PES}(\omega,t_\text{probe})=\nonumber\\
&-i\int dt dt' s(t)s(t')e^{i\omega(t'-t)}G^<(t+t_\text{probe},t'+t_\text{probe})
\end{align}
as a function of $-\omega$ (with some arbitrary rescaling). Here, $s(t)=f_\text{in}(t)$ is the same Gaussian envelope as in the Raman measurements. The photoemission signal also shows an evolution which is symmetric with respect to $t_\text{pump}=15$ and only little indications of heating.  
Consistent with the Raman results, the
plateau feature in this case tracks the Floquet sidebands of the electronic spectrum during the pump. 

\begin{figure}[t]
\begin{center}
\includegraphics[angle=-90, width=0.8\columnwidth]{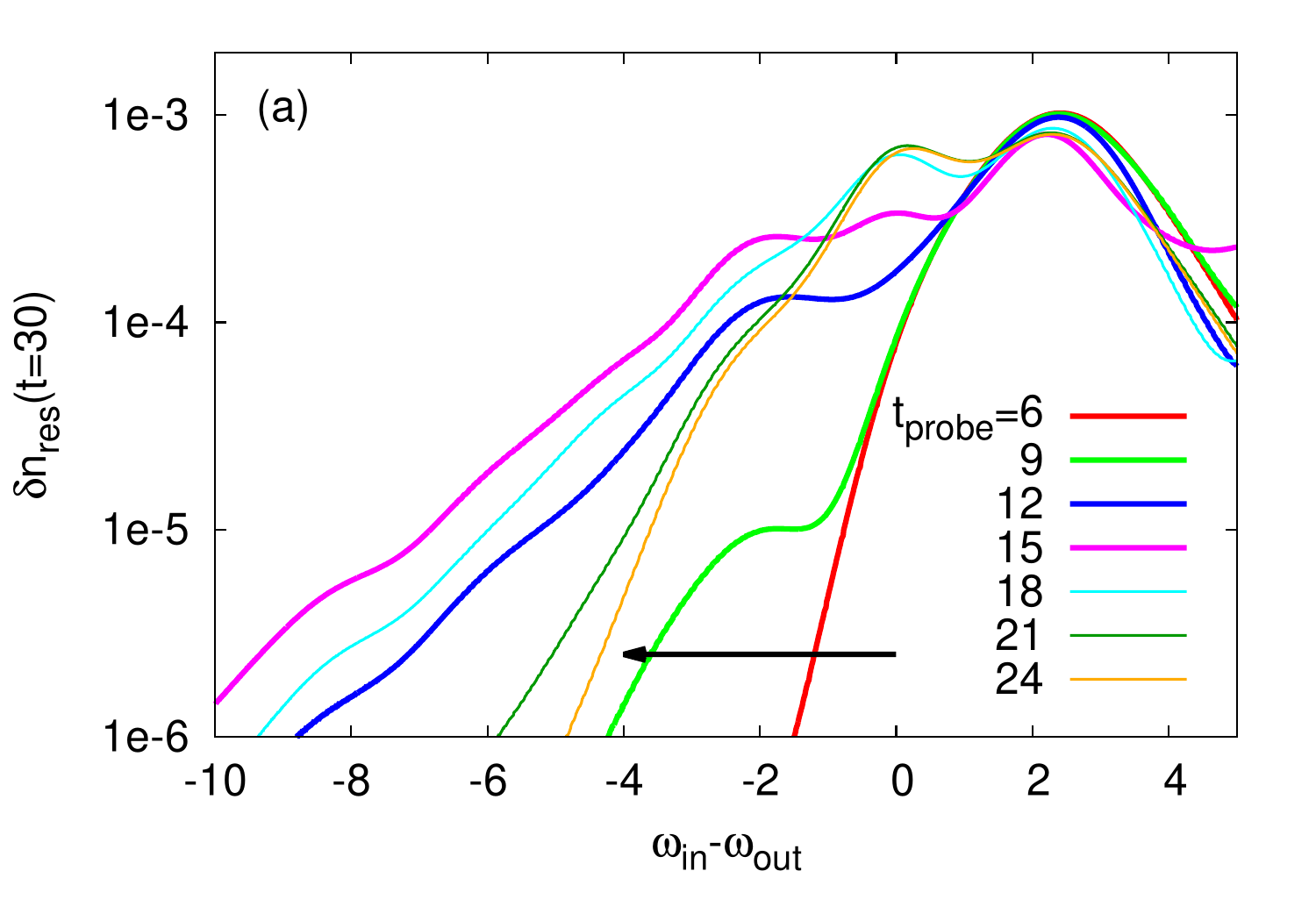}
\includegraphics[angle=-90, width=0.8\columnwidth]{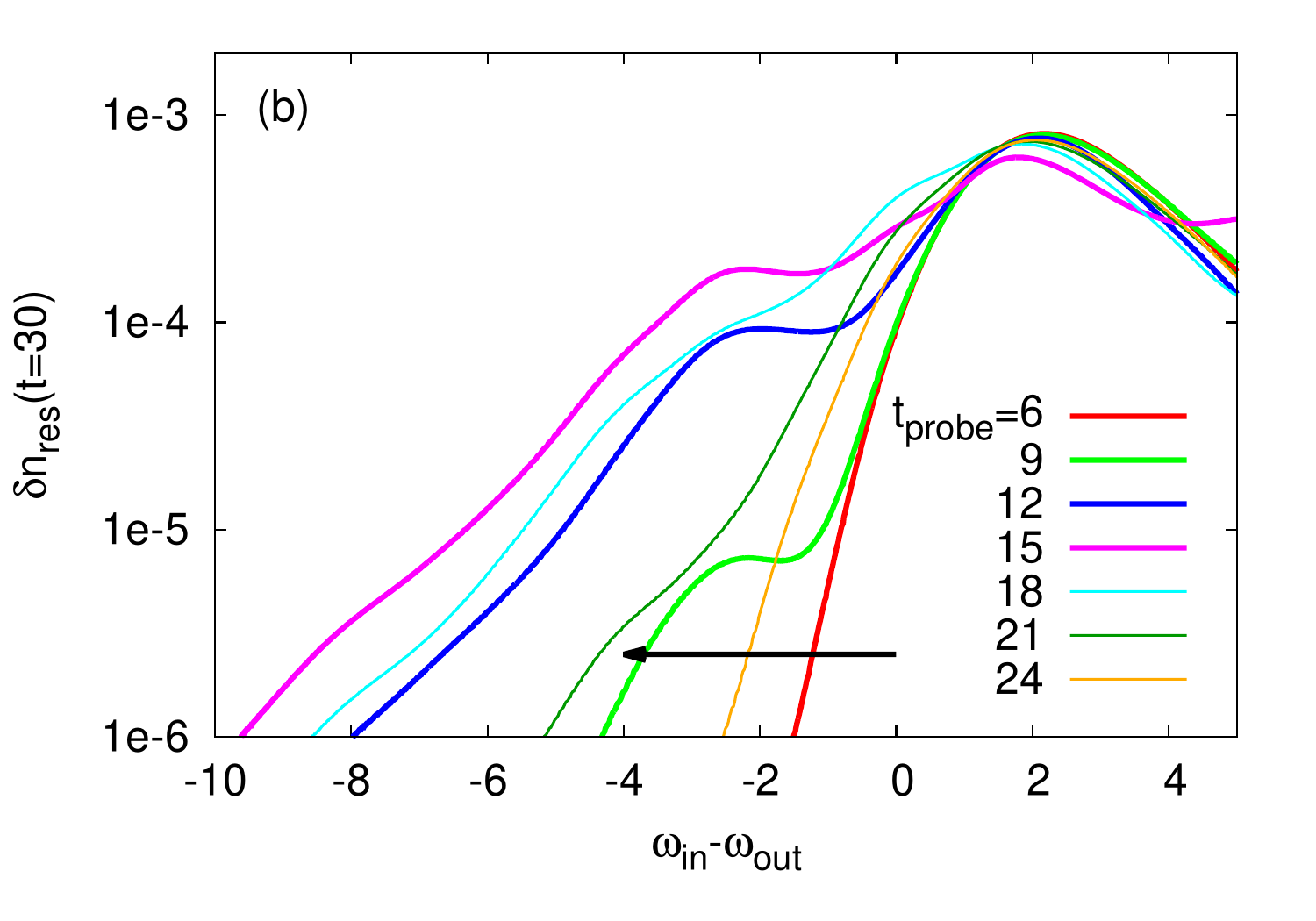}
\caption{
Difference spectra $\delta n_\text{red}=\langle n_\text{res}\rangle_\text{pump+probe}-\langle n_\text{res}\rangle_\text{pump}$ for the pump excited pseudo-gapped metallic systems with $g=0$ (panel (a)) and $g=0.5$ (panel (b)), $U=2$, $\omega_0=1$ and initial $\beta=5$. The pump pulse has amplitude $\mathcal{E}_\text{pump}=4$, frequency $\Omega_\text{pump}=4$ and is centered at $t_\text{pump}=15$. The different lines show the resonant Raman signals obtained for the probe times $t_\text{probe}=6$, 9, \ldots, $24$. Both the pump and the probe pulses have a Gaussian envelope with the same form as shown by the gray shading in Fig.~\ref{fig_eq_nraman_u2_pulse}. The probe pulse has amplitude $\mathcal{E}_\text{in}=0.5$ and frequency $\omega_\text{in}=14$. The black arrow indicates the energy range $-\Omega_\text{pump}\lesssim \omega_\text{in}-\omega_\text{out} \lesssim 0$. 
}
\label{fig_u2_wpump_t_w4}
\end{center}
\end{figure}

Clear signatures of population dynamics can be found in the time-resolved Raman signal of the metallic system if we use a smaller pump frequency $\Omega_\text{pump}$ and weaker phonon coupling $g$. The former results in stronger absorption, and the latter in slower relaxation. Time-resolved resonant Raman difference spectra analogous to Fig.~\ref{fig_u2_wpump_t}(a), but for the lower pump frequency $\Omega_\text{pump}=4$ and weaker electron-phonon coupling are shown in Fig.~\ref{fig_u2_wpump_t_w4}. Panel (a) is for $g=0$ and panel (b) for $g=0.5$ ($U=2$, $\omega_0=1$, initial $\beta=5$). For these parameters, the DMFT solution with the NCA solver yields a pseudo-gapped metal state, with upper and lower Hubbard-band features. The pump pulse with $\Omega_\text{pump}=4$ then leads to a redistribution of charge carriers in these bands. The main difference between $g=0$ and $g=0.5$ is that in the phonon coupled case, the pseudo-gap is less pronounced, and the energy dissipation to the phonons allows the charge carriers to relax to a distribution which is close to that of the initial state, while in the isolated system without phonon coupling, the system will thermalize to a hot-electron distribution. 

This population dynamics is reflected in the Raman spectra. On the one hand, we notice the appearance of a peak at $\omega_\text{in}-\omega_\text{out}\approx 0$, which is associated with charge excitations or deexcitations within the partially populated Hubbard bands. In the $g=0.5$ calculation, this feature appears transiently, while in the $g=0$ case it persists due to the thermalization in the hot electron state. We also notice that in contrast to Fig.~\ref{fig_u2_wpump_t}(a), the evolution of the spectra on the $\omega_\text{in}-\omega_\text{out} < 0$ side is no longer symmetric with respect to the maximum of the pump pulse (at time $t_\text{pump}=15$). This is a clear indication that the signal is not merely due to frequency mixing with the pump pulse, but has also significant anti-Stokes contributions (energy gain from scattering with the nonthermal charge carriers).

\subsubsection{Photo-excited Mott insulating system}

While long-lived nonequilibrium effects are absent in the moderately correlated metallic system, characteristic nonequilibrium properties can be induced by photo-doping a Mott insulating system. In this case the relaxation of photo-doped doublons (doubly occupied sites) and holons (empty sites) leads to the appearance and population of ingap states, as was previously discussed in Ref.~\onlinecite{Werner2015}.   

\begin{figure}[t]
\begin{center}
\includegraphics[angle=-90, width=0.8\columnwidth]{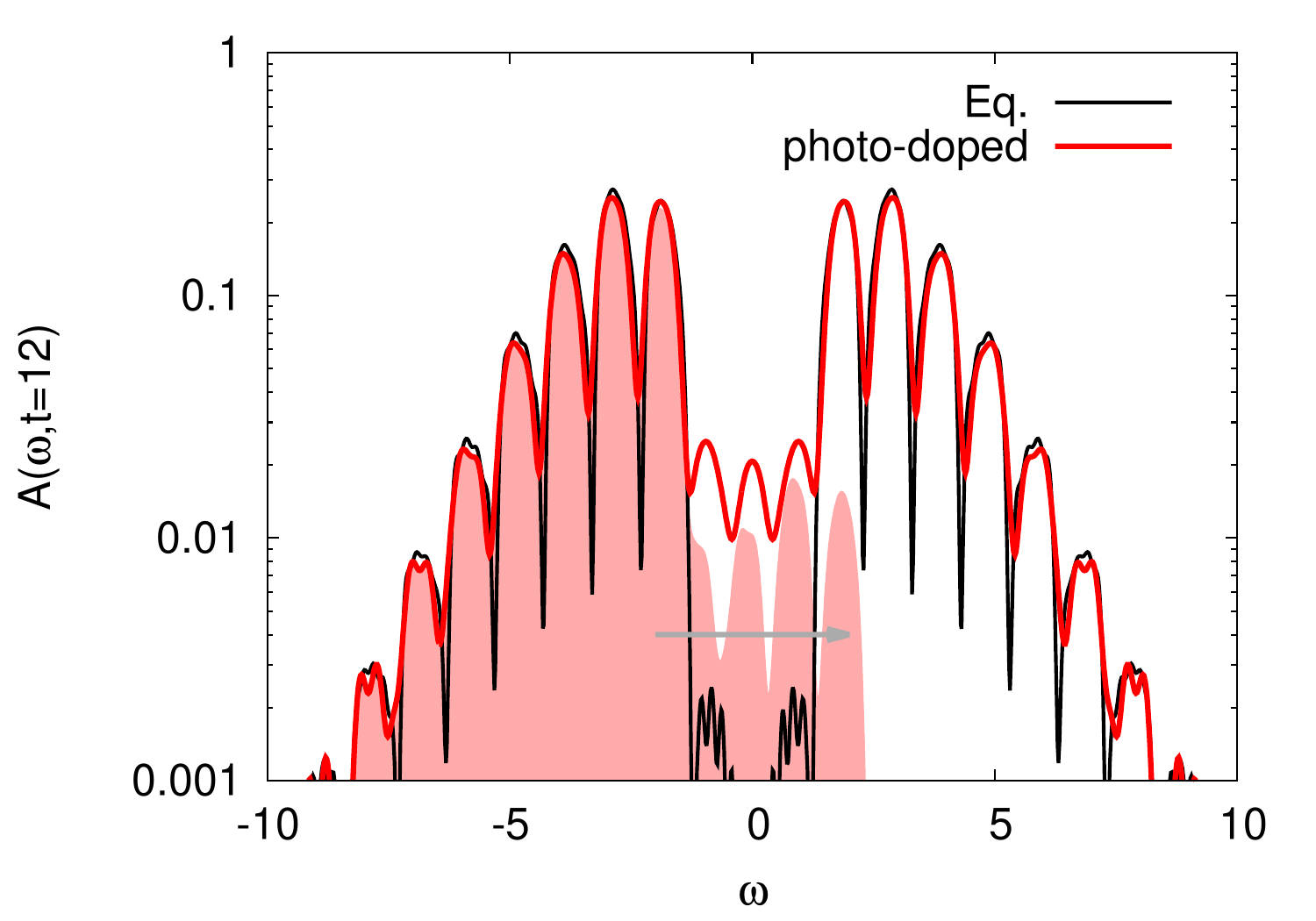}
\caption{DMFT+NCA spectral function of the Holstein-Hubbard model with $U=6$, $g=1$, $\omega_0=1$. The black line shows the equilibrium spectrum for the initial inverse temperature $\beta=5$, while the red line shows the nonequilibrium spectrum at $t=12$, after a photo-doping pulse with $\Omega_\text{pump}=8$ and $\mathcal{E}_\text{pump}=4$. The occupied part of the nonequilibrium spectrum is indicated by the shading. 
}
\label{fig_eq_u6}
\end{center}
\end{figure}

Here, we consider a Mott insulator with $U=6$, $g=1$ and $\omega_0=1$, where a photo-doping pulse with frequency comparable to $U$ induces such long-lived ingap states. While the absorption and hence the weight of these ingap states depends on the pump frequency, the state after the pump exhibits a nonequilibrium distribution which looks qualitatively similar for different pump frequencies, because the photo-doped doublons and holons quickly emit their excess energy to phonons and relax to the first subband of the lower or upper Hubbard band and to the ingap states. The nonequilibrium spectral function and occupation at $t=12$, after a photo-doping excitation with $\Omega_\text{pump}=8$, $\mathcal{E}_\text{pump}=4$, $t_\text{pump}=6$ and the same pump envelope as in the previous section is shown Fig.~\ref{fig_eq_u6}. For comparison, we also plot with a black line the gapped equilibrium spectral function of the initial state.  

\begin{figure}[t]
\begin{center}
\includegraphics[angle=-90, width=0.8\columnwidth]{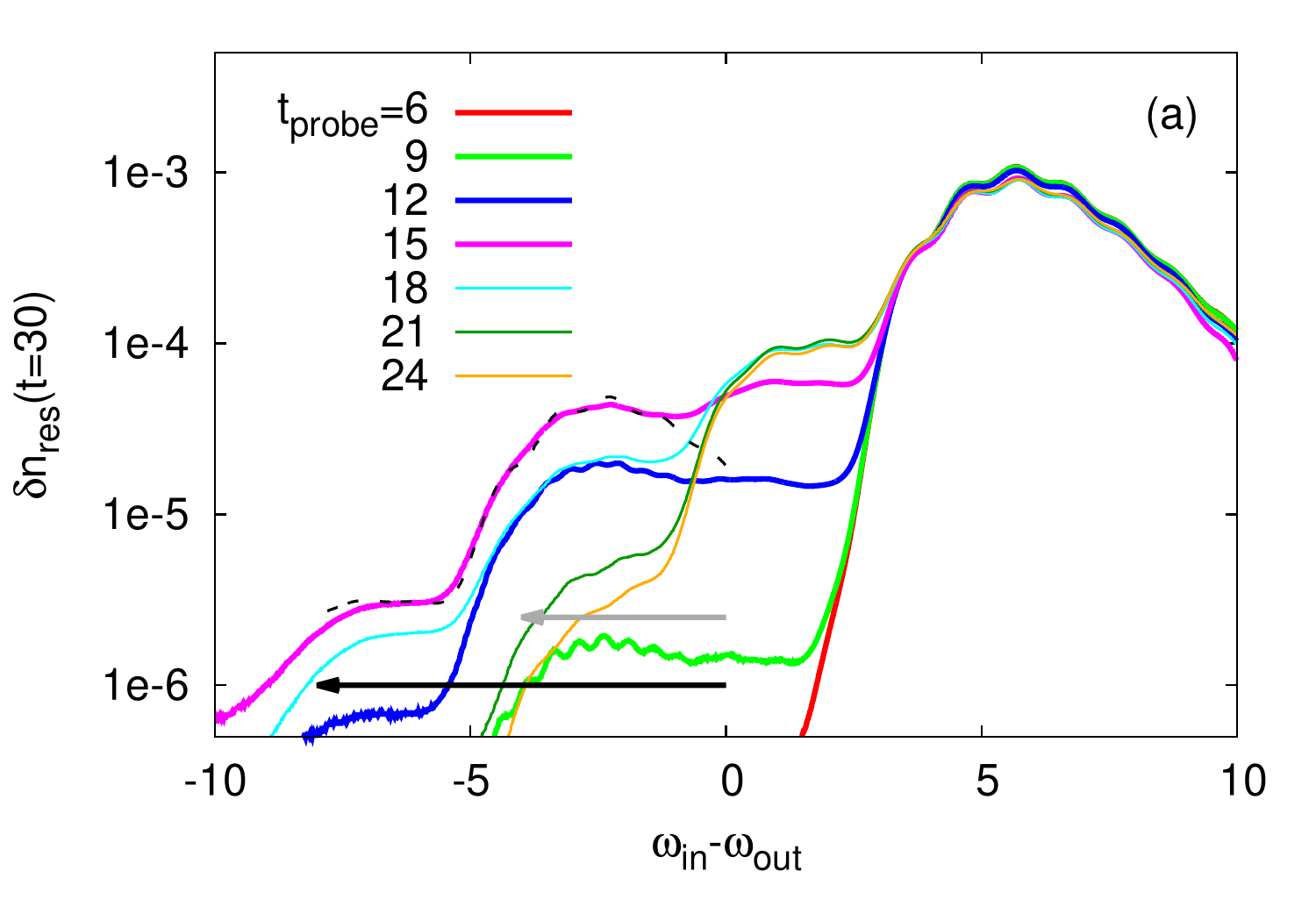}
\includegraphics[angle=-90, width=0.8\columnwidth]{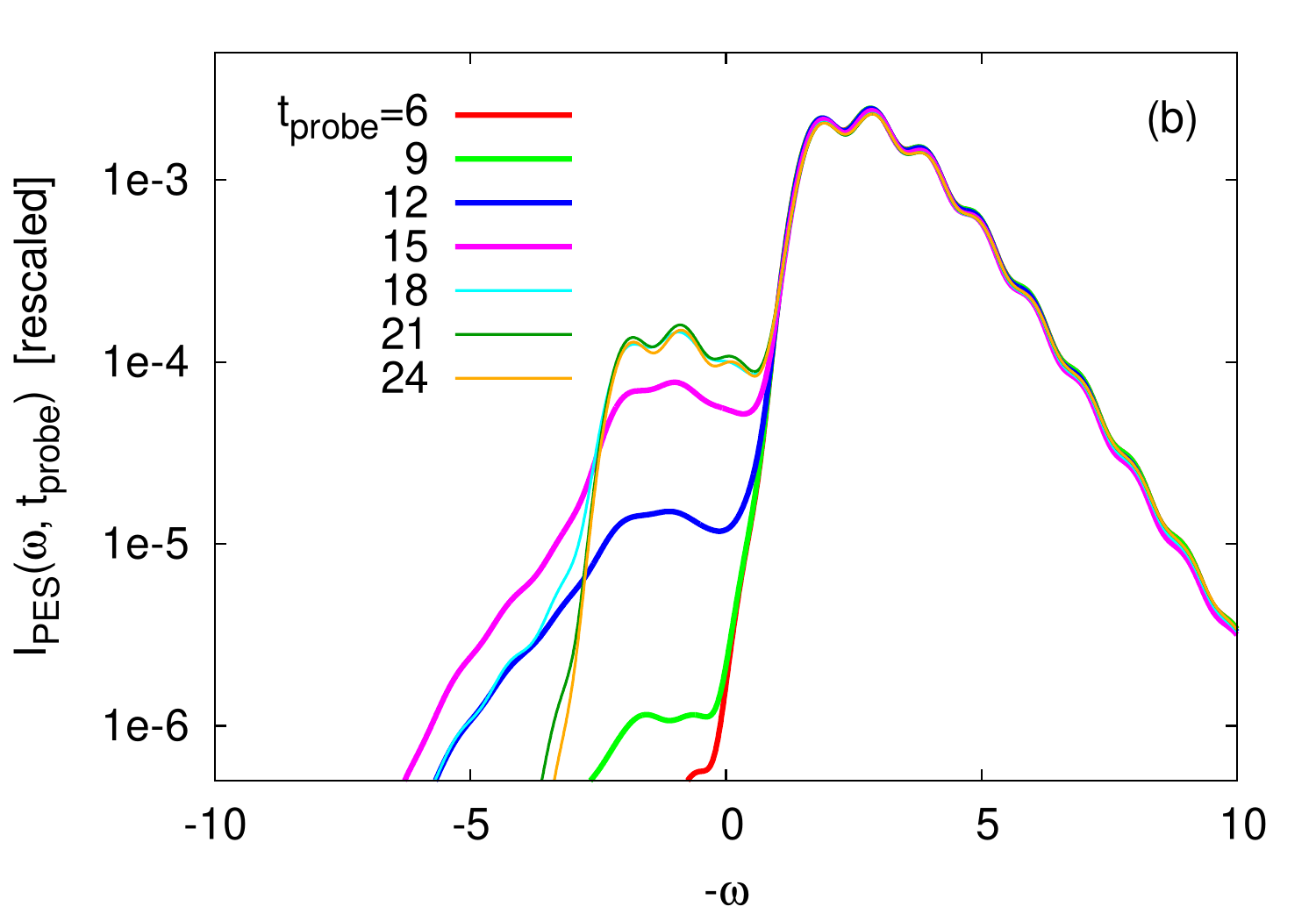}
\caption{Panel (a): Difference spectra $\delta n_\text{red}=\langle n_\text{res}\rangle_\text{pump+probe}-\langle n_\text{res}\rangle_\text{pump}$ for the photo-doped system with $U=6$, $g=1$ and initial $\beta=5$. The photo-doping pulse has frequency $\Omega_\text{pump}=8$, amplitude $\mathcal{E}_\text{pump}=4$ and is centered at $t_\text{pump}=15$. The different lines show the resonant Raman signal obtained for probe pulses at $t_\text{probe}= 6, 9, \ldots, 24$, with amplitude $\mathcal{E}_\text{in}=0.5$ and frequency $\omega_\text{in}=14$. Both the pump and the probe pulses have a Gaussian envelope with the same form as shown by the gray shading in Fig.~\ref{fig_eq_nraman_u2_pulse}. The black arrow indicates the energy range $-\Omega_\text{pump}\lesssim \omega_\text{in}-\omega_\text{out} \lesssim 0$, while the gray arrow shows the range where the long-lived nonthermal population contributes to energy gain (see also Fig.~\ref{fig_eq_u6}). The black dashed line plots a rescaled Stokes signal for $t_\text{probe}=15$, which is shifted down in energy by $\Omega_\text{pump}$. Panel (b): Time-resolved photoemission spectrum $I_\text{PES}(\omega,t_\text{probe})$ as a function of $-\omega$, for the same photodoped state and the same probe envelopes. 
}
\label{fig_u6_wpump_t}
\end{center}
\end{figure}

Because of the long-lived photo-induced ingap states, the Raman spectrum changes qualitatively during and after the photo-doping pulse, and the spectrum does not relax back to a (slightly hotter) equilibrium spectrum on the timescales considered in this study. This is demonstrated in Fig.~\ref{fig_u6_wpump_t}(a), which shows the time evolution of the Raman signal, analogous to Fig.~\ref{fig_u2_wpump_t}(a). Here, the photo-doping pulse is centered at $t_\text{pump}=15$ and the weaker Raman probe pulses are applied at the probe times $t_\text{probe}=6,9,\ldots,24$.  

On the $\omega_\text{in}-\omega_\text{out}>0$ side of the spectrum, we observe the appearance of Stokes peaks in the gap region, which can be associated with the emission of one or two phonons. These peaks remain and even grow beyond $t=15$ (peak amplitude of the pump pulse). This is because the appearance of the photo-induced ingap states and the partial population of the Hubbard bands enables low-energy excitations, and in particular excitations between phonon sidebands with an energy of $n\omega_0$. Since the ingap states and nonthermal populations are long-lived, the photo-induced Stokes peaks remain after the pump. 

On the  $\omega_\text{in}-\omega_\text{out}<0$ side, the photo-doping creates a highly nonthermal plateau structure similar to the one found (during the pump) in the metallic system. 
Some of the transient structures can be associated with frequency mixing, since they have the same shape as the Stokes part of the spectrum (see black dashed line in Fig.~\ref{fig_u6_wpump_t}(a), which shows a shifted and rescaled version of the $t_\text{probe}=15$ Stokes signal) and are present only during the pump. The evolution of the frequency mixing signal is however less symmetric with respect to the peak signal than in the metallic system analyzed in Sec.~\ref{sec_photo_metallic}, because the appearance of the photo-doped ingap states during the pump activates new scattering processes. After the end of the pump pulse, a plateau-like structure remains down to $\omega_\text{in}-\omega_\text{out}\approx -4$, which roughly corresponds to the energy separation between the lower Hubbard band and the highest nonthermally occupied side-band (see gray arrow in Fig.~\ref{fig_eq_u6}). An alternative interpretation is that this edge corresponds to $-(\Omega_\text{pump}-\Delta_\text{gap})$, i.e., to energy gain from relaxation processes within the Hubbard bands.  The former contribution yields a long-lived signal, since the ingap states persist up to the longest simulation times, while the anti-Stokes processes associated with energy gain from the relaxation of electrons or holes within the Hubbard bands exist only during and shortly after the pulse (similar to the metallic case in Fig.~\ref{fig_u2_wpump_t}(a)). 
The short-lived weaker plateau which extends down to $\omega_\text{in}-\omega_\text{out}\approx -8=-\Omega_\text{pump}$ can be interpreted as a frequency mixing signal.

In Fig.~\ref{fig_u6_wpump_t}(b), we plot the (rescaled) photoemission spectrum measured with the same probe pulse envelopes. This spectrum also exhibits a plateau at positive energies associated with the long-lived ingap states (note that the spectrum is plotted as a function of $-\omega$). During the pump, the spectrum also exhibits a tail up to roughly $\omega=\Omega_\text{pump}$, but this tail quickly disappears after the excitation. We thus associate these tails with the population of the Floquet sidebands which are present only during the pump pulse.

\section{Conclusions}
\label{sec:conclusions}

We presented a DMFT-based formalism for the calculation of the resonant Raman scattering signal which does not rely on the evaluation of a four-point correlation function. By explicitly simulating the effect of the incoming light within nonequilibrium DMFT at the classical level, the calculation can be reduced to the evaluation of the lesser component of a two-point correlation function. The approach is not limited to weak probe pulses and for example enables the investigation of hyper Raman scattering. Since it is based on nonequilibrium Green's functions, it can be directly applied to nonequilibrium states created by additional light pulses or other perturbations, and to time-resolved measurements. 

As an application, we evaluated the resonant Raman signal of a metallic and Mott insulating Holstein-Hubbard model with strong electron-phonon coupling. In this case, the spectral function splits into polaron sidebands and the Raman signal exhibits peaks associated with transitions between these sidebands (phonon emission and absorption). In equilibrium, we analyzed the relation between the weights of the Stokes and anti-Stokes signals and showed that the ratio of these weights is consistent with a Boltzmann-type scaling (Eq.~\eqref{eq_thermo}). Hence, in the case of the Holstein-Hubbard model and for the setups considered in this study, Raman-based temperature measurements are reliable even if resonant scattering processes are relevant.   

If the frequency of the incoming light is not resonant to intra-band transitions, and hence the population dynamics plays a minor role, 
the main observation on the $\omega_\text{in}-\omega_\text{out}<0$ side of the spectrum is a frequency-mixing signal during the application of the pump.  
In the case of resonant excitations, photoexcited populations become manifest, e.~g., because the lifetime of the anti-Stokes signal extends beyond the pump pulse envelope by the relaxation time of the nonthermal electrons (which however is rather short in the metallic phase of the Holstein-Hubbard model with strong electron-phonon coupling).

The photo-doping of the Mott system, on the other hand, generates long-lived ingap states and nonthermal electron and phonon populations. These nonthermal populations enable energy gain in the light-scattering process, so that the Raman amplitude for $\omega_\text{in}-\omega_\text{out}<0$ becomes a superposition of the frequency-mixing signal and additional anti-Stokes peaks during the pump. After the pump, it features a long-lived nonthermal plateau with a life-time controlled by the lifetime of the photo-doped doublons/holons and the photo-induced polaronic ingap states.  

In the future, it will be interesting to analyze the relative contributions of the different scattering processes (resonant, nonresonant, mixed) to the Raman signal, and to study the effect of vertex corrections of the type shown in Fig.~\ref{fig_vertex}(c), which are not captured by the presented method. The two-point formalism introduced in this work provides a convenient way for incorporating different classes of vertex corrections, based on physical intuition, such vertex corrections associated with antiferromagnetic or particle-hole scatterings.  

\acknowledgements

We thank D. Fausti, C. Bernhard, Y. Pashkevich, D. Manske and S. Tian for helpful discussions.
The calculations were run on the beo05 cluster at the University of Fribourg. PW acknowledges support from ERC Consolidator Grant No.~724103 and SNSF Grant No.~200021-196966. NT acknowledges support by KAKENHI (Grant No. JP20K03811) and JST FOREST (Grant No. JPMJFR2131).

\appendix

\section{Equivalence of Eqs.~\eqref{eq_raman} and \eqref{dndt}}
\label{app_equivalence}

In the convolutions in Eq.~\eqref{dndt} we replace $B_\gamma$ by the free photon Green's function $B_{0,\gamma}$, whose components are 
\begin{align}
B_{0,\gamma}^<(t,t')&=-i\langle b^\dagger(t')b(t)\rangle_0\nonumber\\
&=-i e^{i\omega_\text{out}(t'-t)}n_B(\omega_\text{out}),\\
B_{0,\gamma}^>(t,t')&=-i\langle b(t) b^\dagger(t')\rangle_0\nonumber\\
&=-ie^{-i\omega_\text{out}(t-t')}(1+n_B(\omega_\text{out})),\\
B_{0,\gamma}^R(t,t')&=-i\langle [b(t),b^\dagger(t')]\rangle_0\nonumber\\
&=-ie^{-i\omega_\text{out}(t-t')}\nonumber\\
&=[B_{0,\gamma}^A(t',t)]^*,\\
B_{0,\gamma}^\lceil(\tau,t')&=-i\langle b(\tau)b^\dagger(t')\rangle_0\nonumber\\
&=-ie^{i\omega_\text{out} t'}e^{-\omega_\text{out}\tau}(1+n_B(\omega_\text{out}))\nonumber\\
&=-[B_{0,\gamma}^\rceil(t',\beta-\tau)]^*,
\end{align}
with $n_B(\omega_\text{out})=1/(e^{\beta\omega_\text{out}}-1)$ the Bose function. 

From the Langreth rules one has $[\Pi*B_{0,\gamma}]^<(t,t')=\int_0^t d\bar t \Pi^R(t,\bar t)B_{0,\gamma}^<(\bar t, t')+\int_0^{t'}d\bar t \Pi^<(t,\bar t)B_{0,\gamma}^A(\bar t,t')-i\int_0^\beta \Pi^\rceil(t,\tau)B_{0,\gamma}^\lceil(\tau,t')$, which gives 
\begin{align}
&(\Pi*B_{0,\gamma}-B_{0,\gamma}*\Pi)^<(t,t')\nonumber\\
&=\int_0^td\bar t\Big[ \Pi^R(t,\bar t)B_{0,\gamma}^<(\bar t, t') -(B_{0,\gamma}^A(\bar t,t'))^*(-\Pi^<(t',\bar t))^*  \Big]\nonumber\\
&+\int_0^{t'}d\bar t\Big[ \Pi^<(t,\bar t)B_{0,\gamma}^A(\bar t, t') -(-B_{0,\gamma}^<(\bar t,t'))^* (\Pi^R(t',\bar t))^*  \Big]\nonumber\\
&-i\int_0^\beta d\tau \Big[ \Pi^\rceil(t,\tau)B_{0,\gamma}^{\lceil}(\tau,t')\nonumber\\
&\hspace{17mm} -(-B^\lceil(\beta-\tau,t))^* (\Pi^\rceil(t',\beta-\tau))^*   \Big].\hspace{10mm}
\end{align}
For $t'=t$ this becomes
\begin{align}
&(\Pi*B_{0,\gamma}-B_{0,\gamma}*\Pi)^<(t,t)\nonumber\\
&=\int_0^{t}d\bar t \, 2\text{Re}\Big[ \Pi^R(t,\bar t)B_{0,\gamma}^<(\bar t, t) + \Pi^<(t,\bar t)B_{0,\gamma}^A(\bar t, t)\Big]\nonumber\\
&\hspace{4mm}-i\int_0^\beta d\tau \, 2\text{Im}\Big[ \Pi^\rceil(t,\tau)B_{0,\gamma}^\lceil(\tau,t) \Big]i\nonumber\\
&= 2\text{Im}\int_{0}^{t}d\bar t e^{i\omega_\text{out}(t-\bar t)}\Big( \Pi^R(t,\bar t)n_B(\omega_\text{out})-\Pi^<(t,\bar t)\Big)\nonumber\\
&\hspace{4mm}-2\text{Re}\int_0^\beta d\tau e^{-\omega_\text{out}\tau}(1+n_B(\omega_\text{out}))e^{i\omega_\text{out}t}\Pi^\rceil(t,\tau). \label{eq_convolution}
\end{align}
We now take the limit $\beta\rightarrow \infty$ ($n_B(\omega_\text{out})\rightarrow 0$) and analytically continue the Matsubara branch to the negative real-time axis: $-i\tau\rightarrow -t'$, $\Pi^\rceil(t,\tau)\rightarrow \Pi^<(t,t')$. This transforms Eq.~\eqref{eq_convolution} to
\begin{align}
&(\Pi*B_{0,\gamma}-B_{0,\gamma}*\Pi)^<(t,t)= -2\text{Im}\int_{0}^{t}d\bar t e^{i\omega_\text{out}(t-\bar t)}\Pi^<(t,\bar t)\nonumber\\
&\hspace{20mm} -2\text{Im}\int_{0}^{\infty}dt' e^{i\omega_\text{out}(t+t')}\Pi^<(t,t').
\end{align}
With the additional substitution $t'=-\bar t$ one finds that Eq.~\eqref{dndt} is equivalent to
\begin{align}
&\frac{d\langle n_\gamma\rangle}{dt} = -2g_\gamma^2\text{Im}\int_{-\infty}^{t}d\bar t e^{i\omega_\text{out}(t-\bar t)}\Pi^<(t,\bar t),
\end{align}
which is consistent with the time derivative of Eq.~\eqref{eq_raman}.

\section{B$_{1g}$ Raman vertex}
\label{app_B1g}

In this Appendix, we show that in the limit $d\rightarrow \infty$, the $B_{1g}$ Raman vertex can be replaced by a constant. 
Here we consider the B$_{1g}$ Raman vertex given by
\begin{align}
M^{B_{1g}}(k)
&\propto
\frac{1}{\sqrt{d}}\sum_\alpha (-1)^\alpha \cos (k_\alpha), 
\end{align}
relevant for nonresonant Raman scattering, but the same line of argument holds for $M^{B_{1g}}(k)\propto \frac{1}{\sqrt{d}}\sum_\alpha (-1)^\alpha \sin (k_\alpha)$, which appears in the calculation of the resonant Raman signal (see main text). 
The system is defined on the $d$-dimensional hypercubic lattice with dispersion
\begin{align}
\varepsilon(k)
&\propto
\frac{1}{\sqrt{d}}\sum_\alpha \cos(k_\alpha).
\end{align}
The Raman response function in the B$_{1g}$ channel can be written as (using a certain function $f(\varepsilon(k))$)
\begin{align}
\label{chi_B1g}
&\chi_{B_{1g}}
=
\sum_k M^{B_{1g}}(k) M^{B_{1g}}(k) f(\varepsilon(k)),
\end{align}
which evaluates to
\begin{align}
&\chi_{B_{1g}}
\propto
\sum_k 
\frac{1}{d}\sum_\alpha (-1)^\alpha \cos (k_\alpha)
\sum_\beta (-1)^\beta \cos (k_\beta)
f(\varepsilon(k))
\notag
\\
&=
\sum_k 
\frac{1}{d}\sum_\alpha \cos^2 (k_\alpha)
f(\varepsilon(k))\nonumber\\
&\hspace{4mm}+\sum_k 
\frac{1}{d}\sum_{\alpha\neq\beta} (-1)^{\alpha+\beta} \cos (k_\alpha)\cos (k_\beta)
f(\varepsilon(k)).
\label{chi B1g}
\end{align}
In the first term, one can replace $\cos^2(k_\alpha)$ with $\frac{1}{2}$ in the large-$d$ limit.
\footnote{This is justified since $\cos^2(k_\alpha)=\frac{1}{2}+\frac{1}{2}\cos(2k_\alpha)$, and $\sum_\alpha \cos(2k_\alpha)$
represents the dispersion with two-site hoppings, which scales as $\sqrt{d}$ in $d\to\infty$.}
To treat the second term, we define
\begin{align}
g_{\alpha\beta}
&=
\sum_k \cos (k_\alpha)\cos (k_\beta) f(\varepsilon(k)),
\end{align}
which for symmetry reasons does not depend on $\alpha, \beta$ when $\alpha\neq\beta$ (i.e., $g_{\alpha\beta}=g$ if $\alpha\neq\beta$).
Then,
\begin{align}
&
\sum_k 
\frac{1}{d}\sum_{\alpha\neq\beta} (-1)^{\alpha+\beta} \cos (k_\alpha)\cos (k_\beta)
f(\varepsilon(k))
\notag
\\
&=
\frac{1}{d}\sum_{\alpha\neq\beta} (-1)^{\alpha+\beta} g_{\alpha\beta}
=
\frac{g}{d}\sum_{\alpha\neq\beta} (-1)^{\alpha+\beta}
\notag
\\
&=
\frac{g}{d}\left[
\left(\sum_{\alpha} (-1)^\alpha\right)^2-\sum_\alpha (-1)^{2\alpha}
\right]
\to
-g 
\quad
(d\to\infty).
\end{align}
Therefore, the second term in Eq.~(\ref{chi B1g}) becomes
\begin{align}
\sum_k 
&\frac{1}{d}\sum_{\alpha\neq\beta} (-1)^{\alpha+\beta} \cos (k_\alpha)\cos (k_\beta)
f(\varepsilon(k))\nonumber\\
&=
-\sum_k \cos(k_\alpha)\cos(k_\beta)f(\varepsilon(k))
\quad
(\alpha\neq\beta)
\notag
\\
&=
-\frac{2}{d(d-1)}\sum_{\alpha<\beta}
\sum_k \cos(k_\alpha)\cos(k_\beta)f(\varepsilon(k))
\notag
\\
&=
-\frac{1}{\sqrt{d(d-1)}}
\sum_k \varepsilon_{\rm fcc}(k) f(\varepsilon(k)),
\end{align}
where we used the dispersion for the $d$-dimensional fcc lattice,
\begin{align}
\varepsilon_{\rm fcc}(k)
&=
\frac{2}{\sqrt{d(d-1)}}\sum_{\alpha<\beta} \cos(k_\alpha)\cos(k_\beta).
\end{align}
Since $\varepsilon_{\rm fcc}(k)$ is related to $\varepsilon(k)$ as \cite{Mueller-Hartmann, Tsuji}
\begin{align}
\varepsilon_{\rm fcc}(k)
&=
(\varepsilon(k))^2-\frac{1}{2}
\end{align}
in the limit of $d\to\infty$, we have
\begin{align}
\sum_k 
&\frac{1}{d}\sum_{\alpha\neq\beta} (-1)^{\alpha+\beta} \cos (k_\alpha)\cos (k_\beta)
f(\varepsilon(k))\nonumber\\
&=
-\frac{1}{\sqrt{d(d-1)}}
\sum_k \left((\varepsilon(k))^2-\frac{1}{2}\right)
f(\varepsilon(k))
\to
0
\end{align}
for $d\to\infty$.
From this it follows that Eq.~(\ref{chi_B1g}) simplifies to
\begin{align}
\chi_{B_{1g}}
\propto
\sum_k f(\varepsilon(k)).
\end{align}
This is proportional the current-current correlation function $\chi_{JJ}$. If the polarization is chosen as $\epsilon=(1,1,1,\ldots)$, the conductivity bubble becomes
\begin{align}
\chi_{JJ}
&\propto
\sum_k \Big( \sum_\alpha \frac{\partial\varepsilon(k)}{\partial k_\alpha}\epsilon_\alpha\Big)\Big(\sum_\beta\frac{\partial\varepsilon(k)}{\partial k_\beta}\epsilon_\beta\Big)
f(\varepsilon(k))
\notag
\\
&=
\sum_k \frac{1}{d} \sum_\alpha \sin^2(k_\alpha)
f(\varepsilon(k))\nonumber\\
&\hspace{4mm}+\sum_k \sum_{\alpha\ne\beta}\frac{1}{d}\sin(k_\alpha)\sin(k_\beta)f(\varepsilon(k))\nonumber\\
&\propto
\sum_k f(\varepsilon(k)),\label{eq_JJ}
\end{align}
where in the middle expression, we replaced $\frac{1}{d} \sum_\alpha \sin^2(k_\alpha)$ by $\tfrac{1}{2}$ and $\frac{1}{d} \sum_\alpha \sin(k_\alpha)$ by zero. 

While the calculations presented in this appendix assume time-independent vertices, analogous results can be derived in the presence of a time-dependent field, by expanding $\sin(k_\alpha-A(t))$ as $\sin(k_\alpha)\cos(A(t))-\cos(k_\alpha)\sin(A(t))$, etc.

\begin{figure}[t]
\begin{center}
\includegraphics[angle=0, width=\columnwidth]{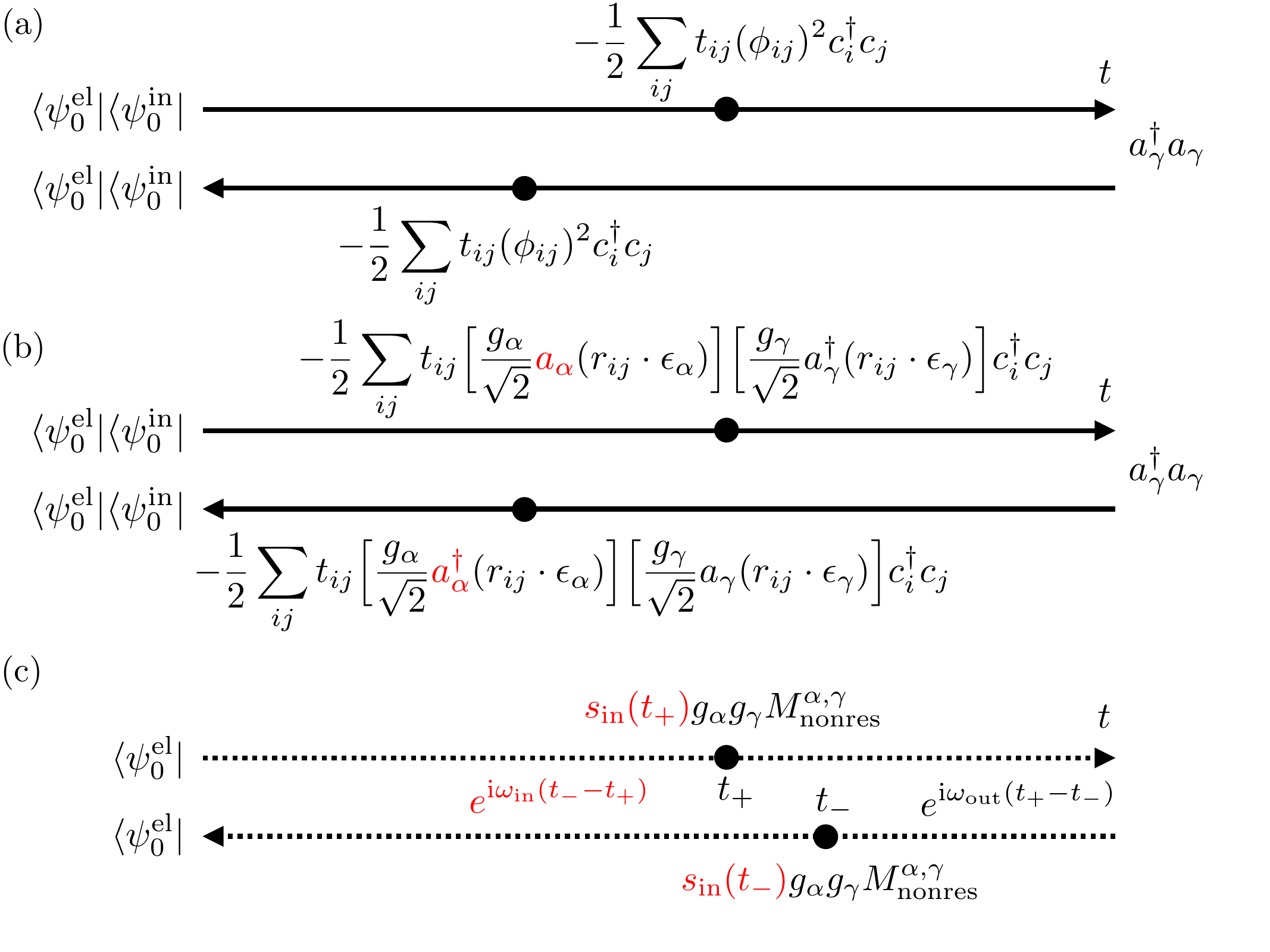}
\caption{
Derivation of the expression for nonresonant Raman scattering. (a) Second order expansion in $\phi_{ij}^2$. Solid arrows represent the time evolution without light-matter coupling, but in the presence of the driving laser field, which populates photon states with index $\alpha\ne\gamma$, where $\gamma$ is the flavor of the outgoing photon. In panel (b), we keep only those photon operators which give a nonzero contribution to the nonresonant process. In panel (c), the driving laser and its effect on the system (red operators) is replaced by a classical field ($a_\alpha\rightarrow s_\text{in}(t)e^{\text{i}\omega_\text{in}t}$), while the expectation value in the photon ground state yields the additional factor $e^{i\omega_\gamma(t_+-t_-)}=e^{i\omega_\text{out}(t_+-t_-)}$.  
}
\label{fig_illustration_nonresonant}
\end{center}
\end{figure}

\begin{figure}[t]
\begin{center}
\hspace{3mm}\includegraphics[angle=-90, width=0.77\columnwidth]{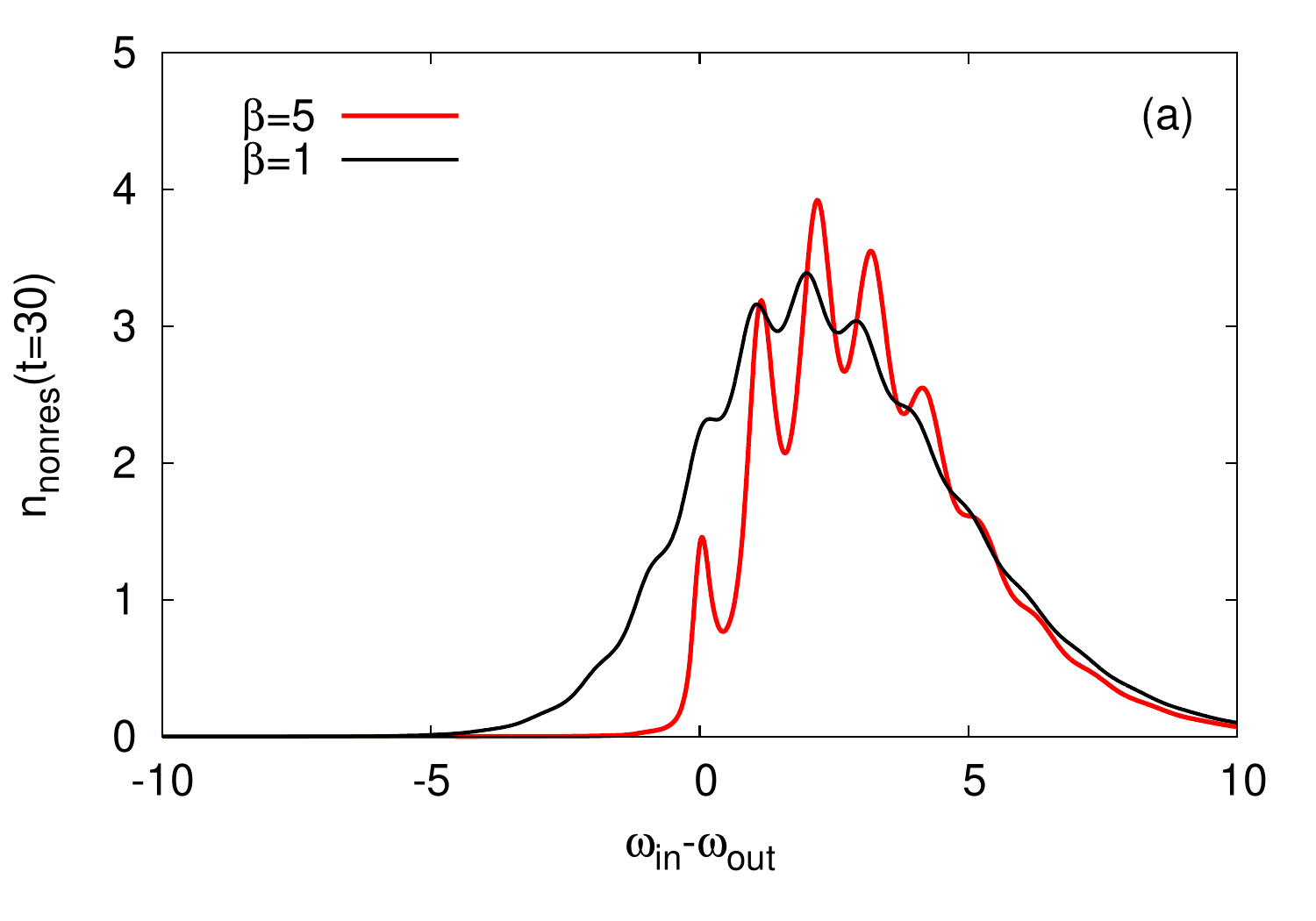}\\
\includegraphics[angle=-90, width=0.8\columnwidth]{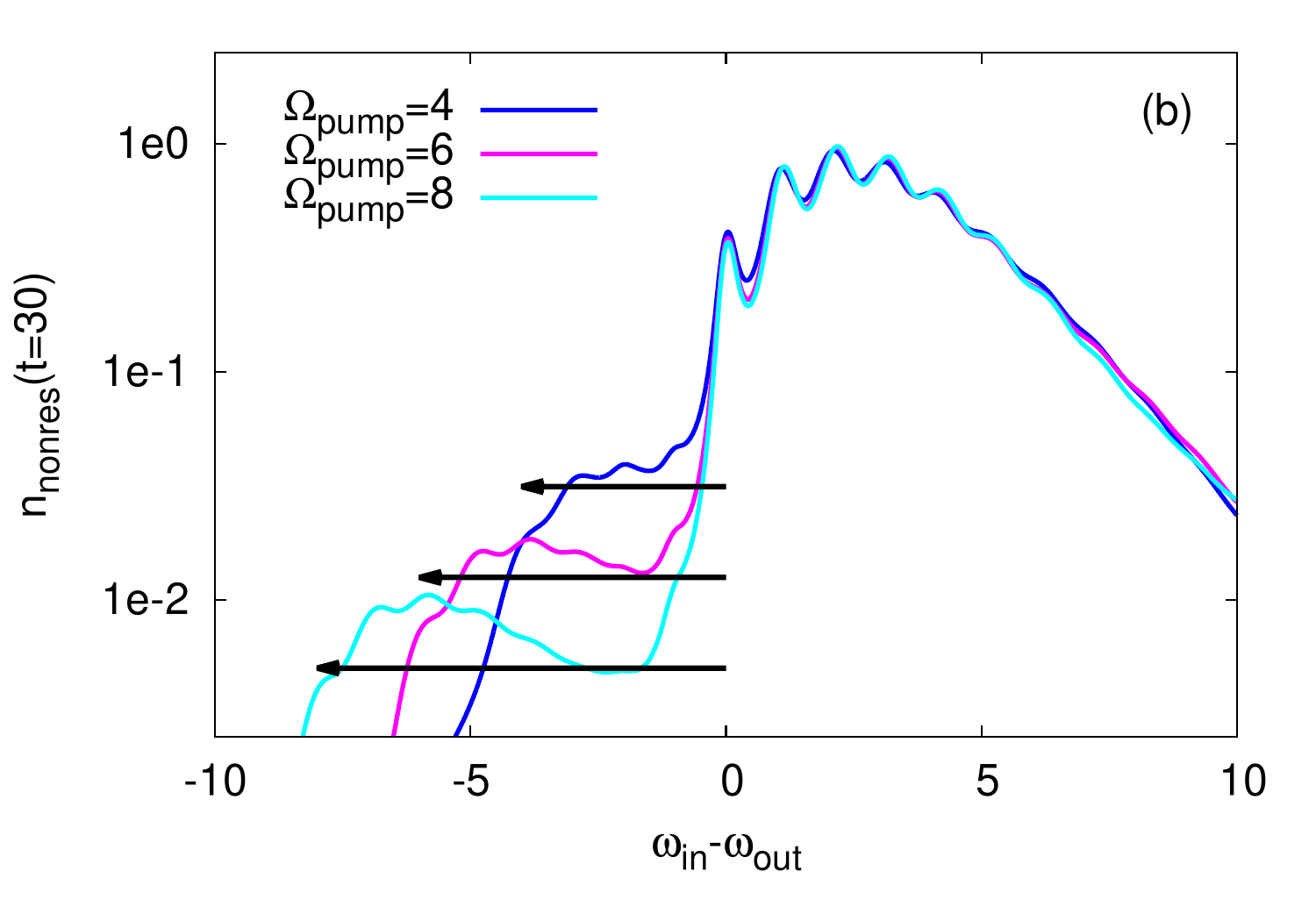}
\caption{
Nonresonant contribution to the Raman signal for $U=2$, $g=1$, $\omega_0=1$ and inverse temperature $\beta=5$ (photon number measured at time $t=30$). 
Panel (a) shows the equilibrium results for $\beta=5$ and $\beta=1$. 
Panel (b) shows nonequilibrium spectra for initial $\beta=5$ and a photo-doping pulse with frequency $\Omega_\text{pump}=4,6,8$ and amplitude $\mathcal{E}_\text{pump}=4$. Black arrows indicate the energy range $-\Omega_\text{pump}\lesssim \omega_\text{in}-\omega_\text{out} \lesssim 0$. 
The Raman probe pulse has amplitude $\mathcal{E}_\text{in}=0.5$ and frequency $\omega_\text{in}=14$.
}
\label{fig_nonresonant}
\end{center}
\end{figure}

\section{Nonresonant Raman diagram}
\label{app_nonresonant}

The nonresonant scattering diagram can be obtained by inserting the $\sim (\phi_{ij})^2$ term in the expansion (\ref{eq_expansion}) on the forward and backward time contour, as shown in Fig.~\ref{fig_illustration_nonresonant}(a). We may then replace the incoming (outgoing) photon on the forward (backward) branch by a classical field, using the substitutions
\begin{equation}
a_\alpha \rightarrow s_\text{in}(t)e^{\text{i}\omega_\text{in}t},\quad a^\dagger_\alpha \rightarrow s_\text{in}(t)e^{-\text{i}\omega_\text{in}t}
\end{equation}
with $s_\text{in}(t)$ the envelope of the light pulse. The photon expectation value in the ground state can then be evaluated and leads to a factor $e^{i\omega_\text{out}(t_+-t_-)}$, as in the main text. The sums over the lattice sites in the vertices can be transformed into sums over momenta, which yields 
\begin{align}
M_\text{nonres}^{\alpha,\gamma}&=-\frac{1}{2}\sum_k M^{\alpha,\gamma}_k n_k,\label{vertex_nonres}\\
M^{\alpha,\gamma}_k&=\frac{1}{2}\sum_{\mu,\nu}\epsilon_{\alpha,\mu} \frac{\partial^2 \varepsilon_k}{\partial {k_\mu}\partial {k_\nu}}\epsilon_{\gamma,\nu}.
\end{align} 
Here $\varepsilon_k$ is the dispersion of the lattice, while $\epsilon_{\alpha(\gamma)}$ is the polarization of the incoming (outgoing) light. We thus obtain 
\begin{align}
&\langle n_\text{nonres}^{\gamma}\rangle (t,\omega_\text{in}-\omega_\text{out})  \nonumber\\
&= i g_\alpha^2 g_\gamma^2\int_{-\infty}^t dt_+ dt_- s_\text{in}(t_+)s_\text{in}(t_-) e^{\text{i}(\omega_\text{out}-\omega_\text{in})(t_+-t_-)}\nonumber\\
&\hspace{35mm}\times \Pi^<_\text{nonres}(t_+,t_-)\nonumber\\
&= -g_\alpha^2 g_\gamma^2 \int_{-\infty}^t dt_+ dt_- s_\text{in}(t_+)s_\text{in}(t_-) \text{Im}\Big[e^{\text{i}(\omega_\text{out}-\omega_\text{in})(t_+-t_-)}\nonumber\\
&\hspace{35mm}\times \Pi^<_\text{nonres}(t_+,t_-)\Big],\label{eq_nonresraman}\\
&\Pi_\text{nonres}(t_+,t_-) = -i \langle M_\text{nonres}^{\alpha,\gamma}(t_+)M_\text{nonres}^{\alpha,\gamma}(t_-)\rangle\nonumber\\
&\hspace{22.5mm}\approx \frac{i}{4}\sum_{k\sigma} M^{\alpha,\gamma}_k(t_+) G_{k\sigma}(t_+,t_-) \nonumber\\
&\hspace{35mm}\times G_{k\sigma}(t_-,t_+) M^{\alpha,\gamma}_k(t_-).\label{eq_pi_nonresraman}
\end{align}

Apart from the pulse envelope factors, this corresponds to the standard expression for the nonresonant Raman signal.\cite{Devereaux2007} A possible nonequilibrium state of the system, induced for example by a pump pulse, enters through the nonequilibrium lattice Green's functions $G_{k\sigma}$. 
In contrast to the procedure discussed for the resonant Raman signal in the main text, we do not explicitly simulate here the nonequilibrium state produced by the incoming light of the Raman measurement, and hence the ``Fourier integral" involves the energy difference $\Delta\omega=\omega_\text{in}-\omega_\text{out}$, which can be positive or negative.

Up to a rescaling factor, the results obtained from Eqs.~(\ref{eq_nonresraman}) and (\ref{eq_pi_nonresraman}) are very similar to those presented in the main text for the resonant scattering amplitude, both for the equilibrium and photo-excited systems. In Fig.~\ref{fig_nonresonant} we show data analogous to Fig.~\ref{fig_eq_nraman_u2}(b) and to Fig.~\ref{fig_u2_wpump}, for $\alpha=\gamma$ in the vertex \eqref{vertex_nonres} and $g_\alpha g_\gamma/2=1$. The model parameters and the pump excitation are the same as in the main text.

\section{Relation to the optical conductivity}

The correlation function \eqref{eq_conn} appearing in our two-point formalism looks like a conductivity. However, inferring a direct connection between the {\it resonant} Raman signal and the conductivity from this observation is not possible, because Eq.~\eqref{eq_conn} is the correlation function of the driven system (driven by the Raman probe, which enters the propagators in this expression), while the conductivity is determined by the undriven current-current correlation function. For the {\it nonresonant} contribution, on the other hand, an explicit connection to the conductivity can be made in $d=\infty$.

The optical conductivity is given by\cite{Aoki2014}
\begin{align}
\sigma(t,t')
&=
-\int_{t'}^t d\bar t\, \chi^R(t,\bar t),
\end{align}
where in DMFT
\begin{align}
&\chi^R(t,t')
=
-2\sum_{k\sigma} \gamma_k\, {\rm Im}[G_{k\sigma}^R(t,t')G^<_{k\sigma}(t',t)] \gamma_k
\notag
\\
&\hspace{3mm}=
i\sum_{k\sigma} \gamma_k [G_{k\sigma}^R(t,t')G_{k\sigma}^<(t',t)+G_{k\sigma}^<(t,t')G_{k\sigma}^A(t',t)] \gamma_k
\end{align}
and $\gamma_k=\sum_\alpha \frac{\partial\varepsilon_k}{\partial k_\alpha}\epsilon_\alpha$. In the limit of $d\to\infty$ and for polarization $\epsilon=(1,1,1,\ldots)$ 
one can replace $\gamma_k^2$ with $\frac{1}{2}$ (see derivation of Eq.~\eqref{eq_JJ}).

The Fourier transform of $\sigma(t,t')$ is 
\begin{align}
\sigma(\omega)
&=
\int_{-\infty}^t d\bar t\, e^{i\omega(t-\bar t)}\sigma(t,\bar t)
\notag
\\
&=
-\int_{-\infty}^t d\bar t\, e^{i\omega(t-\bar t)}
\int_{\bar t}^t d\bar t'\, \chi^R(t,\bar t'), 
\end{align}
and hence
\begin{align}
i\omega\sigma(\omega)
&=
\int_{-\infty}^t d\bar t\, [\partial_{\bar t}e^{i\omega(t-\bar t)}]
\int_{\bar t}^t d\bar t'\, \chi^R(t,\bar t')
\notag
\\
&=
\int_{-\infty}^t d\bar t\, e^{i\omega(t-\bar t)}\chi^R(t,\bar t)\\
&=
i\int_{-\infty}^t d\bar t\, e^{i\omega(t-\bar t)}
\sum_{k\sigma} \gamma_k \big[G_{k\sigma}^>(t,\bar t)G_{k\sigma}^<(\bar t,t)\nonumber\\
&\hspace{27mm}-G_{k\sigma}^<(t,\bar t)G_{k\sigma}^>(\bar t,t)\big]\gamma_k.
\end{align}
Using $[G^>(t,t')]^\ast=-G^>(t',t)$ and $[G^<(t,t')]^\ast=-G^<(t',t)$ one finds
\begin{align}
&{\rm Re}\,[\omega\sigma(\omega)]
=
\frac{1}{2}\int_{-\infty}^t d\bar t\, e^{i\omega(t-\bar t)}
\sum_{k\sigma} \gamma_k \big[G_{k\sigma}^>(t,\bar t)G_{k\sigma}^<(\bar t,t)\nonumber\\
&\hspace{44mm}-G_{k\sigma}^<(t,\bar t)G_{k\sigma}^>(\bar t,t)\big]\gamma_k
\notag
\\
&\hspace{5mm}
+\frac{1}{2}\int_{-\infty}^t d\bar t\, e^{-i\omega(t-\bar t)}
\sum_{k\sigma} \gamma_k \big[G_{k\sigma}^>(\bar t,t)G_{k\sigma}^<(t,\bar t)\nonumber\\
&\hspace{35mm}-G_{k\sigma}^<(\bar t,t)G_{k\sigma}^>(t,\bar t)\big]\gamma_k.
\label{optical conductivity}
\end{align}

On the other hand, it follows from Eqs.~\eqref{eq_nonresraman} and \eqref{eq_pi_nonresraman} that the nonresonant contribution of the Raman signal 
can be expressed as ($\Delta\omega=\omega_\text{in}-\omega_\text{out}$)
\begin{align}
&\frac{d}{dt} \langle n_\gamma^{\rm nonres}\rangle(t)\\
&=
-2g_\alpha^2 g_\gamma^2 s(t){\rm Im}\int_{-\infty}^t d\bar t\, e^{-i\Delta\omega(t-\bar t)}s(\bar t)\Pi_{\rm nonres}^<(t,\bar t)\nonumber\\
&=
g_\alpha^2 g_\gamma^2 s(t) \int_{-\infty}^t d\bar t\, s(\bar t)e^{-i\Delta\omega(t-\bar t)}i\Pi_{\rm nonres}^<(t,\bar t)\nonumber\\
&\hspace{4mm}+g_\alpha^2 g_\gamma^2 s(t)\int_{-\infty}^t d\bar t\, s(\bar t)e^{i\Delta\omega(t-\bar t)}\big[i\Pi_{\rm nonres}^<(t,\bar t)\big]^\ast,
\end{align}
where
\begin{align}
i\Pi_{\rm nonres}^<(t,t')
&=
-\frac{1}{4}\sum_{k\sigma} M_k^{\alpha,\gamma}G_{k\sigma}^<(t,t')G_{k\sigma}^>(t',t) M_k^{\alpha,\gamma}.
\end{align}
Substituting this into the above equation, and assuming a slowly varying envelope, we obtain
\begin{align}
&\frac{d}{dt} \langle n_\gamma^{\rm nonres}\rangle(t)\nonumber\\
&
\propto \int_{-\infty}^t d\bar t\, e^{-i\Delta\omega(t-\bar t)}
\sum_{k\sigma} M_k^{\alpha,\gamma}G_{k\sigma}^<(t,\bar t)G_{k\sigma}^>(\bar t,t)M_k^{\alpha,\gamma}
\notag
\\
&
\hspace{4mm}+\int_0^t d\bar t\, e^{i\Delta\omega(t-\bar t)}
\sum_{k\sigma} M_k^{\alpha,\gamma}G_{k\sigma}^<(\bar t,t)G_{k\sigma}^>(t,\bar t)M_k^{\alpha,\gamma}.
\label{nonresonant Raman}
\end{align}

By comparing Eqs.~(\ref{optical conductivity}) and (\ref{nonresonant Raman}), 
and taking into account that the difference
between $\gamma_k$ and $M_k^{\alpha,\gamma}$ becomes negligible 
in $d\to\infty$ (the average of $\sin^2(k)$ and $\cos^2(k)$ becomes $\tfrac 12$),
we see that the optical conductivity is, up to prefactors, the 
difference
of the Raman signals with $\Delta\omega=\omega$ and $\Delta\omega=-\omega$:
\begin{align}
{\rm Re}\, [\omega\sigma(\omega)]
&\propto
\frac{d}{dt} \langle n_\gamma^{\rm nonres}\rangle(t)\bigg|_{\Delta\omega=\omega}
-\frac{d}{dt} \langle n_\gamma^{\rm nonres}\rangle(t)\bigg|_{\Delta\omega=-\omega}.
\end{align}

The derivations presented in this appendix are valid for time-independent vertices, but they can be generalized to electric field driven systems 
by expanding  $\sin(k_\alpha-A(t))$ as $\sin(k_\alpha)\cos(A(t))-\cos(k_\alpha)\sin(A(t))$, etc.


\begin{thebibliography}{99}
\bibitem{Giannetti2016} C. Giannetti, M. Capone, D. Fausti, M. Fabrizio, and F. Parmigiani, Advances in Physics {\bf 65}, 58 (2016). 
\bibitem{Torre2021} A.~de la Torre, D.~M.~Kennes, M.~Claassen, S.~Gerber, J.~W.~McIver, and M.~A.~Sentef, Rev. Mod. Phys. {\bf 93}, 041002 (2021)
\bibitem{Sobota2021} J. A. Sobota, Y. He, and Z.-X. Shen, Rev. Mod. Phys. {\bf 93}, 025006 (2021).
\bibitem{Stamm2010} C. Stamm, N. Pontius, T. Kachel, M. Wietstruk, and H. A. D\"urr, Phys. Rev. B {\bf 81}, 104425 (2010).
\bibitem{Baykusheva2022} D.~R.~Baykusheva, H.~Jang, A.~A.~Husain, S.~Lee, S.~F.~.R.~TenHuisen, P.~Zhou, S.~Park, H.~Kim, J.~Kim, H.~Kim, M.~Kim, S.~Park, P.~Abbamonte, B.~J.~Kim, G.~D.~Gu, Y.~Wang, and M.~Mitrano, Phys. Rev. X {\bf 12}, 011013 (2022).
\bibitem{Lojewski2023} T. Lojewski {\it et al.}, 
Mater. Res. Lett. {\bf 11}, 655 (2023).
\bibitem{Yang2017} J.-A. Yang, S. Parham, D. Dessau, and D. Reznik, Sci. Rep. {\bf 7}, 40876 (2017).
\bibitem{Dean2016} M. Dean et al., Nature Mat. {\bf 15}, 601 (2016).
\bibitem{Wang2020} M.~Mitrano  and Y.~Wang, Comm.~Phys.~{\bf 3}, 184 (2020).
\bibitem{Ament2011} L. J. P. Ament, M. van Veenendaal, T. P. Devereaux, J. P. Hill, and J. van den Brink. Rev. Mod. Phys. {\bf 83}, 705 (2011).
\bibitem{Devereaux2007} T. P. Devereaux and Rudi Hackl, Rev. Mod. Phys. {\bf 79}, 175 (2007).
\bibitem{Saichu2009} R. P. Saichu, I. Mahns, A. Goos, S. Binder, P. May, S. G. Singer, B. Schulz, A. Rusydi, J. Unterhinninghofen, D. Manske, P. Guptasarma, M.~S. Williamsen, and M. R\"ubhausen, Phys. Rev. Lett. {\bf 102}, 177004 (2009).
\bibitem{Fausti2009} D.~Fausti, O.~V.~Misochko, and P.~H.~M.~van~Loosdrecht, Phys. Rev. B {\bf 80}, 161207(R) (2009).
\bibitem{Kang2010} K.~Kang, D.~Abdula, D.~G.~Cahill, and M.~Shim, Phys. Rev. B {\bf 81}, 165405 (2010).
\bibitem{Yan2009} H.~Yan, D.~Song, K.~F.~Mak, I.~Chatzakis, J.~Maultzsch, and T.~F.~Heinz, Phys. Rev. B {\bf 80}, 121403(R) (2009).
\bibitem{Matveev2019} O. P. Matveev, A. M. Shvaika, T. P. Devereaux, and J. K. Freericks, Phys. Rev. Lett. {\bf 122}, 247402 (2019).
\bibitem{Wang2018} Y.~Wang, T.~P.~Devereaux, and C.~Chen, Phys. Rev. B {\bf 98}, 245106 (2018).
\bibitem{Walter2020} M.~Walter and M.~Moseler, J. Chem. Theory Comput. {\bf 16}, 576 (2020).
\bibitem{Freericks2001} J. Freericks and T. Devereaux, Phys. Rev. B {\bf 64}, 125110 (2001).
\bibitem{Shvaika2004} A. M. Shvaika, O. Vorobyov, J. K. Freericks, and T. P. Devereaux, Phys. Rev. Lett. {\bf 93}, 137402 (2004).
\bibitem{Shvaika2005} A. M. Shvaika, O. Vorobyov, J. K. Freericks, and T. P. Devereaux, Phys. Rev. B {\bf 71}, 045120 (2005).
\bibitem{Matveev2010} O. P. Matveev, A. M. Shvaika, and J. K. Freericks, Phys. Rev. B {\bf 82}, 155115 (2010).
\bibitem{Chen2019} Y. Chen, Y. Wang, C. Jia, B. Moritz, A. M. Shvaika, J. K. Freericks, and T. P. Devereaux, Phys. Rev. B {\bf 99}, 104306 (2019).
\bibitem{Shao2016} C. Shao, T. Tohyama, H.-G. Luo, and H. Lu, Phys. Rev. B {\bf 93}, 195144 (2016).
\bibitem{Werner2019} P. Werner, J. Li, D. Golez, and M. Eckstein, Phys. Rev. B {\bf 100}, 155130 (2019).
\bibitem{Zawadzki2020} K. Zawadzki, L. Yang, and A. E. Feiguin, Phys. Rev. B {\bf 102}, 235141 (2020).
\bibitem{Rincon2021} J. Rincon and A. E. Feiguin, Phys. Rev. B {\bf 104}, 085122 (2021).
\bibitem{Eckstein2021} M. Eckstein and P. Werner, Phys. Rev. B {\bf 103}, 115136 (2021).
\bibitem{Werner2021a} P. Werner, S. Johnston, and M. Eckstein, Europhys. Lett. {\bf 133}, 57005 (2021).
\bibitem{Georges1996} A. Georges, G. Kotliar, W. Krauth, and M. J. Rozenberg, Rev. Mod. Phys. {\bf 130}, 68 (1996).
\bibitem{Hariki2018} A. Hariki, M. Winder, and J. Kunes, Phys. Rev. Lett. {\bf 121}, 126403 (2018).
\bibitem{Hariki2020} A. Hariki, M. Winder, T. Uozumi, and J. Kunes, Phys. Rev. B {\bf 101}, 115130 (2020).
\bibitem{Werner2021b} P. Werner and M. Eckstein, Phys. Rev. B {\bf 104}, 085155 (2021).
\bibitem{Werner2023} P. Werner, F. Petocchi, and Martin Eckstein, Phys. Rev. B {\bf 107}, 035157 (2023).
\bibitem{Peierls1933} R. Peierls, Z. Physik {\bf 80}, 763 (1933). 
\bibitem{Aoki2014} H. Aoki, N. Tsuji, M. Eckstein, M. Kollar, T. Oka, and P. Werner, Rev. Mod. Phys. {\bf 86}, 779 (2014).
\bibitem{Stefanucci_book} G. Stefanucci and R. van Leeuwen, Nonequilibrium Many-Body Theory of Quantum Systems: A Modern Introduction, (Cambridge University Press, Cambridge, 2013).
\bibitem{Nessi} M. Sch\"uler, D. Golez, Y. Murakami, N. Bittner, A. Herrmann, H. Strand, P. Werner, and M. Eckstein, Computer Physics Communications {\bf 257}, 107484 (2020).
\bibitem{Li2020} J.~Li, D.~Golez, G.~Mazza, A.~J.~Millis, A.~Georges, and M.~Eckstein, Phys. Rev. B {\bf 101}, 205140 (2020).
\bibitem{Kurana1990} A. Khurana, Phys. Rev. Lett. 64, (1990).
\bibitem{Eckstein2008} M. Eckstein and M. Kollar, Phys. Rev. B {\bf 78}, 205119 (2008).
\bibitem{Tsuji2009} N. Tsuji, T. Oka and H. Aoki, Phys. Rev. Lett. {\bf 103}, 047403 (2009).
\bibitem{Turkowski2005} V. Turkowski and J. K. Freericks, Phys. Rev. B {\bf 77}, 205102 (2005).
\bibitem{Eckstein2010} M. Eckstein and P. Werner, Phys. Rev. B {\bf 82}, 115115 (2010).
\bibitem{Werner2013} P. Werner and M. Eckstein, Phys. Rev. B {\bf 88}, 165108 (2013).
\bibitem{Werner2015} P. Werner and M. Eckstein, Europhys. Lett. {\bf 109}, 37002 (2015).
\bibitem{Sentef2013} M. Sentef, A. F. Kemper, B. Moritz, J. K. Freericks, Z.-X. Shen, and T. P. Devereaux, Phys. Rev. X {\bf 3}, 041033 (2013).
\bibitem{Murakami2015} Y. Murakami, P. Werner, N. Tsuji, and H. Aoki, Phys. Rev. B {\bf 91}, 045128 (2015).
\bibitem{Madzharova2017} F. Madzharova, Z. Heinera, and J. Kneipp, Chem. Soc. Rev. {\bf 46}, 3980 (2017). 
\bibitem{Murakami2015} Y. Murakami, P. Werner, N. Tsuji, and H. Aoki, Phys. Rev. B {\bf 91}, 045128 (2015).
\bibitem{Freericks2009} J. K. Freericks, H. R. Krishnamurthy, and T. Pruschke, Phys. Rev. Lett. {\bf 102}, 136401 (2009).
\bibitem{Kauch2020} A. Kauch, P. Pudleiner, K. Astleithner, P. Thunstr\"om, T. Ribic, and K. Held, Phys. Rev. Lett. {\bf 124}, 047401 (2020).
\bibitem{Simard2021} O. Simard, S. Takayoshi, and P. Werner, Phys. Rev. B {\bf 103}, 104415 (2021).
\bibitem{Worm2021} P. Worm, C. Watzenb\"ock, M. Pickem, A. Kauch, and K. Held, Phys. Rev. B {\bf 104}, 115153 (2021).
\bibitem{Simard2021b} O. Simard, M. Eckstein, and P. Werner, Phys. Rev. B {\bf 104}, 245127 (2021).
\bibitem{Aslamazov1968} L. G. Aslamazov and A. Larkin, Sov. Phys. Solid State {\bf 10}, 87 (1968).
\bibitem{Canovi2014} E. Canovi, P. Werner, and M. Eckstein, Phys. Rev. Lett. {\bf 113}, 265702 (2014).
\bibitem{Mueller-Hartmann} E. M\"uller-Hartmann, Z. Phys. B: Condens. Matter {\bf 74}, 507 (1989).
\bibitem{Tsuji} N. Tsuji, T. Oka, H. Aoki, Phys. Rev. B {\bf 78}, 235124 (2008).

\end{thebibliography}
\end{document}